\documentclass[acmsmall,screen,nonacm]{acmart}
\usepackage[T1]{fontenc}
\newif\ifdraft\draftfalse

\usepackage[capitalize]{cleveref}

\usepackage{enumitem}

\usepackage{amsmath}

\usepackage{tikz}
\usetikzlibrary{positioning,calc}

\pgfdeclarelayer{background}
\pgfsetlayers{background,main}

\colorlet{col/archi-sec-link}{ACMPurple}
\colorlet{col/OXS}{MediumBlue}
\colorlet{col/UXS}{OrangeRed}

\tikzstyle{colOXS} = [draw=col/OXS, fill=col/OXS!30]
\tikzstyle{colUXS} = [draw=col/UXS, fill=col/UXS!30]

\ifdraft
    \paperwidth=\dimexpr\paperwidth + 6cm\relax
    \makeatletter%
    \@mparswitchfalse%
    \makeatother%
    \usepackage[tickmarkheight=2pt,backgroundcolor=orange!75]{todonotes}
\else
    \usepackage[disable]{todonotes}
\fi

\newcommand\comm[2][]{\todo[linecolor=Plum,backgroundcolor=Plum!25,bordercolor=Plum,#1]{#2}}

\usepackage{listings}

\lstdefinestyle{common}
{
    columns=flexible,
    gobble=4,
    breaklines=true,
    basicstyle=\small\tt,
}

\lstdefinestyle{ocaml}
{
    style=common,
    language={[Objective]Caml},
}

\lstdefinestyle{smtlib}
{
    style=common,
    language={Lisp},
    keywordstyle=\color{DarkViolet},
    commentstyle=\color{Grey},
    alsoletter={<=>+},
    deletekeywords={cons,some,length,append},
    morekeywords={
        lambda,
        List,
        forall, len, map, fold, fold_left, filter, value,
        declare-const, define-const,
        declare-datatype, declare-datatypes,
        declare-fun, define-fun, define-fun-rec,
        set-logic,
        check-sat,
        <=,+,*,=,
        ite, match, with,
        true,false,
    },
}

\lstnewenvironment{SMT-LIB}
{\lstset{style=smtlib}}
{}

\lstnewenvironment{OCaml}
{\lstset{style=ocaml}}
{}

\newcommand{\smtlib}[1]{\lstinline[style=smtlib]{#1}}
\newcommand{\ml}[1]{\hbox{\lstinline[style=ocaml]{#1}}} %

\usepackage{xspace}

\newcommand\evalNumFilesDuelist{39\xspace}
\newcommand\evalNumFilesSat{23\xspace}
\newcommand\evalNumFilesUnsat{16\xspace}
\newcommand\evalNumFilesGenerated{\hbox{2,256}\xspace}
\newcommand\evalNumFilesParam{752\xspace}
\newcommand\evalNumLinesDuelist{739\xspace}
\newcommand\evalNumLinesGenerated{\hbox{56,315}\xspace}

\newcommand\UXS{Under-approximating solver\xspace}
\newcommand\OXS{Over-approximating solver\xspace}
\newcommand\uxs{under-approximating solver\xspace}
\newcommand\oxs{over-approximating solver\xspace}
\newcommand\duelist{\mbox{DueList}\xspace}

\newcommand\eg{\emph{e.g.}\xspace}

\usepackage{mathtools}
\usepackage{mathpartir}

\newcommand{\Rule}[1]{\hyperlink{#1}{\TirName {#1}}}

\newcommand\sym[1]{\textit{#1}} %

\newcommand\conc[1]{\textbf{#1}} %

\newcommand\li[2]{\ensuremath{{#1}_{\conc{#2}}}} %
\newcommand\liln[2]{\ensuremath{\li{#1}1, \ldots, \li{#1}{#2}}}

\newcommand\model[1]{\ensuremath{\mathcal{#1}}}
\newcommand\modelSMT{\ensuremath{\mathcal{M}_\mathsf{SMT}}}
\newcommand\modelAccess[2]{\ensuremath{{#1}[{#2}]}}

\newcommand\comb[1]{\smtlib{#1}}
\newcommand\len{\comb{len}\xspace}
\newcommand\forallc{\comb{forall}\xspace}
\newcommand\map{\comb{map}\xspace}
\newcommand\fold{\comb{fold}\xspace}
\newcommand\filter{\comb{filter}\xspace}
\newcommand\ite{\smtlib{ite}\xspace}
\newcommand\match{\smtlib{match}\xspace}

\newcommand\funspace{{\;}}
\newcommand\app[2]{#1\funspace#2}
\newcommand\apptwo[3]{#1\funspace#2\funspace#3}
\newcommand\appthree[4]{#1\funspace#2\funspace#3\funspace#4}

\newcommand\applen[1]{\app\len{#1}}
\newcommand\appmap[2]{\apptwo\map{#1}{#2}}
\newcommand\appfilter[2]{\apptwo\filter{#1}{#2}}
\newcommand\appfold[3]{\appthree\fold{#1}{#2}{#3}}
\newcommand\appite[3]{\appthree\ite{#1}{#2}{#3}}
\newcommand\appmatch[1]{\apptwo\match{#1}\smtlib{with}}
\newcommand\applambda[2]{\lambda {#1}.\;#2}
\newcommand\applambdatwo[3]{\lambda {#1}\;{#2}.\;#3}

\newcommand\listtype[1]{\smtlib{List}~{#1}}

\newcommand\result[1]{\textsf{\textbf{#1}}} %
\newcommand\mkSAT[1]{\ensuremath{\result{SAT}_{\result{#1}}}\xspace}
\newcommand\mkUNSAT[1]{\ensuremath{\result{UNSAT}_{\result{#1}}}\xspace}
\newcommand\mkUNKNOWN[1]{\ensuremath{\result{UNKNOWN}_{\result{#1}}}\xspace}
\newcommand\SAT{\mkSAT{}}
\newcommand\UNSAT{\mkUNSAT{}}
\newcommand\UNKNOWN{\mkUNKNOWN{}}
\newcommand\SMTSAT{\mkSAT{SMT}}
\newcommand\SMTUNSAT{\mkUNSAT{SMT}}
\newcommand\SMTUNKNOWN{\mkUNKNOWN{SMT}}

\newcommand\lenvar[1]{|#1|}

\newcommand\extractLen[2]{#1 \to #2}
\newcommand\rewriteFixed[2]{#1 \hookrightarrow #2}

\newcommand\compLengthSolver{Orchestrator}

\newcommand\compRewriteBreak{Deforestation}
\newcommand\compFixedLength{List encoder}

\newcommand\bench[1]{\emph{#1}}

\usepackage{array}
\newcolumntype{H}{>{\setbox0=\hbox\bgroup}c<{\egroup}@{}}

\begin{abstract}
  Formal verification tools commonly rely on SMT solvers to automatically reason
  about programs, leveraging a range of logical theories, e.g., linear integer
  arithmetic, arrays, or strings, to encode program constructs and verification conditions.
  Despite recent advances, such solvers still struggle when reasoning about recursive
  data structures such as lists, which are pervasive in modern functional languages.
  Additionally, lists are commonly used in conjunction with higher-order combinators to, e.g.,
  generically apply a function to all elements of the list.

  In this work, we provide first-class support for reasoning
  about lists within SMT solvers. We focus on lists of arbitrary size that, following
  the map-reduce paradigm, can be manipulated exclusively through a set of abstract combinators.
  To this end, we introduce \duelist, an abstraction-refinement approach
  geared towards list reasoning, which we implement on top of off-the-shelf SMT solvers.
  To evaluate the efficiency of our approach, we assemble a diverse set of
  \evalNumFilesParam benchmarks
  curated from previous works and real-world programs, and compare \duelist
  against state-of-the-art solvers such as Z3 and CVC5. Our experimental evaluation
  shows that \duelist extends reasoning facilities of existing solvers, allowing
  to conclude about the (un)satisfiability of a larger range of problems, while
  outperforming existing solvers in the vast majority of previously supported cases.

\end{abstract}

\author{Pierre Goutagny}
\orcid{0000-0003-0876-7188}
\affiliation{\institution{Univ. Lille, Inria, CNRS, Centrale Lille, UMR 9189 CRIStAL}\city{F-59000 Lille}\country{France}}
\email{pierre.goutagny@inria.fr}

\author{Aymeric Fromherz}
\orcid{0000-0003-2642-543X}
\affiliation{\institution{Inria}\city{Paris}\country{France}}

\author{Raphaël Monat}
\orcid{0000-0001-8487-0326}
\affiliation{\institution{Univ. Lille, Inria, CNRS, Centrale Lille, UMR 9189 CRIStAL}\city{F-59000 Lille}\country{France}}

\title{\duelist: A Theory of Lists with Combinators for SMT Solvers}

\begin{document}
\maketitle

\section{Introduction}

SMT solvers~\cite{z3,cvc5} are the backbone of many modern program verification tools.
By providing theories to reason about diverse program features such as arithmetic, but also
strings, arrays, and other data structures~\cite{DBLP:conf/cav/BjornerN24,DBLP:journals/fmsd/LiangRTTBD16,DBLP:conf/issta/PerryMZC17,DBLP:conf/fmcad/KomuravelliBGM15,DBLP:journals/jar/PhamGW16}, they allow to reason about the safety and
correctness of programs, and underpin widely-used symbolic execution and deductive verification
tools~\cite{DBLP:conf/osdi/CadarDE08,pasareanu2010symbolic,DBLP:conf/vstte/LeinoM10,DBLP:conf/esop/FilliatreP13,DBLP:conf/popl/SwamyHKRDFBFSKZ16}
with applications to real-world,
industrial use cases~\cite{DBLP:conf/cav/Rungta22,DBLP:conf/icse/KimKKJ12, everest2026TOPLAS}.

Despite their expressiveness, SMT solvers struggle to automatically reason about widespread datatypes such as functional lists.
Issues are twofold. First, such datatypes, and functions operating on them,
are typically inductively defined, which remains challenging for SMT solvers to handle automatically~\cite{sabharwal2024cvc5induction}
despite recent advances~\cite{DBLP:conf/vmcai/ReynoldsK15}.
Second, these datatypes are generally manipulated using generic, higher-order functions to, e.g.,
modify all elements of a list according to a user-provided function by applying a \map combinator, or to ensure that a given
predicate holds for all elements in a collection through the use of a generic \forallc function.
To address this issue, several works proposed to leverage user-provided annotations, or \emph{contracts}, to reason about arbitrarily
complex functions at the cost of full automation~\cite{DBLP:conf/esop/FilliatreP13,leino2010dafny,DBLP:conf/popl/SwamyHKRDFBFSKZ16,tobin2012higher}.
Other approaches~\cite{torlakLightweightSymbolicVirtual2014} restrict reasoning to lists with a concrete, fixed size, thus enabling the SMT encoding of a symbolic list
as a set of symbolic variables corresponding to its elements, but forgoing arbitrary length reasoning.
Previous works have also focused on establishing decision procedures for generic algebraic datatypes~\cite{DBLP:conf/cade/ReynoldsB15,DBLP:conf/vmcai/ReynoldsK15} which are
implemented in mainstream SMT solvers, but experimental results show that the approach is not tailored enough to scale (\Cref{sec:eval}).

In this work, we instead consider a different setting, where we wish to reason on symbolic, arbitrarily sized lists, but restrict which operations can be performed.
Our key observation, following the map-reduce paradigm, is that many classes of programs treat collections as abstract types, and only operate on them through a set of generic combinators,
without ever requiring pattern-matching on the underlying representation.
As an example, consider
the program in \Cref{fig:running.ml}, inspired by a real-world implementation
computing social benefits~\cite{DBLP:journals/pacmpl/MerigouxCP21}.
This program computes whether a parent meets sufficient conditions in terms of child custody to be eligible for family benefits. %
The child custody score is computed as a sum over all children in the household (here encoded as a \fold) of the part they contribute to, which depends on whether custody is exclusive, shared, or not awarded (obtained through \map).

\begin{figure}%
    \centering
    \begin{tabular}{c}
\begin{lstlisting}[style=ocaml]
let custody_score =
  List.fold (+) 0
    (List.map (fun child -> match child.custody with
         | Sole -> 2
         | Shared -> 1
         | None -> 0)
        children)
in (custody_score > custody_score_threshold)
&& List.length children > child_count_threshold
\end{lstlisting}
    \end{tabular}
    \caption{OCaml running example, inspired from real-world family benefits computation}%
    \Description{OCaml running example} %
    \label{fig:running.ml}
\end{figure}

To reason about abstract lists with combinators, we therefore propose a novel approach based on an abstraction-refinement methodology,
through a cooperation between over-approximating and under-approximating encodings of
quantifier-free list constraints
building on top of modern, off-the-shelf SMT solvers.
We provide an overview of our approach in \Cref{sec:architecture}.
The over-approximating solver allows us to conclude about the unsatisfiability
of a set of constraints.
In contrast, the under-approximating solver is able to
generate models pertaining to the original problem when satisfiable, or to refine the search space considered by the over-approximating
encoding when unsuccessful.
Each of these solvers relies on list abstractions, including length abstractions (\Cref{sec:length}) as well
as abstractions pertaining to the shape of the symbolic lists (\Cref{sec:fold}), complemented with list-specific
simplification rules and common SMT optimizations (\Cref{sec:optimizations}).
We implement this approach in a prototype tool, called \duelist.

To evaluate our methodology,
we curate a set of parametric benchmarks
capturing a range of programs using lists with combinators.
\comm{P: @Aymeric j'ai essayé de partir de ta suggestion}
To cover a wide range of problem sizes,
we generate \evalNumFilesParam unique benchmarks
by making the constraints on list lengths range
over several orders of magnitude (\Cref{sec:eval}).
To this end, we first propose an extension of the SMT-LIB language
to encode list values, as well as the combinators we consider in this work, namely, \map, \fold, and \len (\Cref{sec:language}).
We then manually translate into this language a collection of test cases extracted from prior works on list reasoning~\cite{DBLP:journals/jar/PhamGW16} and functional
synthesis~\cite{DBLP:journals/corr/abs-1711-11438,DBLP:journals/cacm/AlurSFS18}, as well as examples from the OCaml and Catala compiler test suites~\cite{leroy2024ocaml,DBLP:journals/pacmpl/MerigouxCP21}.
Using this benchmarking suite, we perform an ablation study to evaluate
the usefulness of different heuristics and optimizations implemented as part of our approach.
We implement an automated translation from our list theory to other SMT formats
requiring universal quantification:
the datatype and array theories of SMT-LIB~\cite{SMT-LIBv2.7},
and the sequence theory of Z3~\cite{z3}.
We leverage this automated translation to compare the performance and expressiveness of our prototype to other state-of-the-art tools.
\duelist{} shows good scalability in our parametric benchmarks and outperforms other solvers in the vast majority of the cases.

\paragraph{Data-Availability Statement}
Both \duelist and the benchmarks presented in this paper will be available under open source licenses.
To foster reproducibility, we also plan to submit an artifact, consisting in a Docker container including scripts
to reproduce the experimental evaluation described in this paper.
Finally, we plan to submit the SMT-LIB problems
that \duelist generated from our benchmarks (\Cref{sec:eval:comp})
to the SMT-COMP competition to encourage further support for lists
in state-of-the-art solvers.
Our benchmarks and SMT-LIB translations are available to reviewers
as supplementary material.

\section{Extending SMT-LIB with abstract lists and combinators}
\label{sec:language}

We start this paper by describing our target language.
To foster reproducibility and facilitate interaction with existing work, we propose
an extension of the SMT-LIB language~\cite{SMT-LIBv2.7} to integrate lists
with combinators.

\paragraph{Lists as algebraic datatypes}%
In most functional programming languages,
lists are defined as an abstract datatype with two constructors,
\smtlib{nil} for the empty list,
and \smtlib{cons}, which takes a head element and a tail list and returns a new list combining the two.
To operate on lists, users typically define operators and functions recursively, by destructing
list values through pattern matching and reconstructing them using \smtlib{nil} and \smtlib{cons}.

While this approach is highly versatile and enables programmers to implement
a range of higher-order functions, it significantly hinders symbolic reasoning over lists.
Indeed, to work around SMT solvers' limitations related to inductive reasoning, SMT queries
pertaining to lists are commonly encoded as a disjunction of cases, where one
considers fixed-size lists of different lengths and specializes queries accordingly to unfold
recursive operations~\cite{torlakLightweightSymbolicVirtual2014}. For arbitrary functions, even reasoning specifically about the lengths of lists
can prove tricky: performing pattern matching on lists or constructing a new one impacts their lengths,
and such operations can be combined arbitrarily to, e.g., append two lists or filter elements according to a user-provided predicate,
leading again to inductive reasoning on a recursively defined \len function.

In this paper, we contend that the \smtlib{nil}/\smtlib{cons} approach
is too general for many use cases and code patterns.
Drawing inspiration from the map-reduce approach and from typical collection library
APIs, we thus propose to focus on a restricted version of abstract
lists with generic combinators.

\paragraph{Abstract lists with combinators}
In our theory,
lists are purely abstract values
that are neither constructed from, nor breakable into,
a head and tail.
Rather, we expose standard, generic combinators that directly apply to list values,
namely, \len, \map, and \fold.
$\applen l$ represents the length of $l$, described symbolically given that lists can have an arbitrary size.
$\appmap fl$ represents a list
whose elements are the output of $f$
applied to elements of $l$.
$\appfold fil$ represents the reduction of $l$ to a scalar
by chaining calls to $f$ on the elements of $l$,
starting with $i$.
It is arguably the most expressive combinator in our language,
as it can represent the kind of operations enabled by recursion,
that is,
ones that combine elements of a list while keeping an accumulator.

Importantly, these combinators are treated as opaque, primitive operations, and therefore
are not defined through standard recursive functions.
In particular, our theory and encodings are quantifier-free.
As we will show in \Cref{sec:length}, this encoding allows us to reason on list lengths without reasoning on their shape or content.
Thanks to our support of non-recursive datatypes (including tuples and option types), these generic combinators can be used to implement a variety of standard combinators,
including \comb{exists}, \comb{for_all}, \comb{sum}, or \comb{contains}.
We present in \Cref{sec:eval:corpus} the set of functions implemented by our corpus of benchmarks.

\paragraph{An SMT-LIB extension for lists.}
To facilitate the development, use, and comparison of SMT solvers, SMT-LIB provides a common language equipped with a well-defined semantics to encode a variety of problems,
and is the \textit{de facto} entry point to interact with modern solvers. In order to provide a foundation for benchmarking suites operating on lists, we propose an extension
of a subset of the SMT-LIB language~\cite{SMT-LIBv2.7}.
Our language leverages existing SMT-LIB support for integer and boolean sorts and operations, as well as
non-recursive function definitions, non-recursive datatypes, and lambda expressions.

We extend this subset with a native datatype for lists, called \smtlib{List}, which is
parameterized by the type (or \smtlib{Sort} in SMT parlance) of the list elements.
Using standard notation, a symbolic variable $\sym l$ representing a list of integers
can be declared by:
\begin{SMT-LIB}
    (declare-const l (List Int))
\end{SMT-LIB}

Then, we add a keyword for lists of known, concrete length,
that are defined explicitly with all of their elements.
We pass the sort of the list to this constructor
to statically type empty lists.
\begin{SMT-LIB}
    (List.value Int (0 (+ x y) 3)) ; [0, x + y, 3], with x and y symbolic integers
    (List.value Bool ()) ; []
\end{SMT-LIB}
Since we focus on problems where lists appear as parameters,
and where \duelist is tasked with creating a satisfying model containing concrete lists,
this syntax appears more often in the output of \duelist
than in its input files.

As a last step, we finally expose list combinators as special functions in the language.
To define functions used in \map and \fold combinators,
we leverage the support and syntax for lambda-abstractions
defined in the SMT-LIB standard since its version 2.7~\cite{SMT-LIBv2.7}.
\begin{SMT-LIB}
    (List.len l)
    (List.map (lambda ((x Int)) (<= x 10)) l)
    (List.fold (lambda ((acc Int) (x Int)) (+ acc x)) 0 l)
\end{SMT-LIB}

To simplify the use of our language extension with existing solvers,
we also provide encodings into standard SMT theories, such
as arrays or datatypes. We describe this in more detail in \Cref{sec:eval}.

\paragraph{Notation used in this paper}
We use the SMT-LIB extension presented in this section
both in our implementation of \duelist and to encode
our benchmarking suite focusing on list problems (\Cref{sec:eval}).
To lighten the presentation, we however rely in the rest of this
paper on an ML-like syntax with infix operators, as for instance
shown in \Cref{fig:running.ml}.

\section{\duelist: An Overview}
\label{sec:architecture}

Before delving into the technical details of our approach, we now provide a high-level overview of \duelist.
We first describe its architecture in \Cref{sec:architecture:description} before illustrating its application to our running example in \Cref{sec:architecture:example}.

\subsection{Architecture}
\label{sec:architecture:description}

\duelist operates on top of existing SMT solvers. It extends reasoning capabilities with a theory of lists,
whose elements can be of any type supported by the underlying solver.
Typically, this includes the integer,
boolean, and datatypes theories of SMT-LIB, supported by mainstream solvers such as Z3~\cite{z3} or CVC5~\cite{cvc5}.

The core solving procedure of \duelist consists of
an abstraction-refinement loop, a common architecture for
extending SMT solvers to novel theories and application
domains~\cite{
    cimatti2018incremental,
    bauer2010don,
    pulina2010abstraction,
    DBLP:conf/fmcad/SinhaSMSW12,
    DBLP:journals/sttt/BryantKOSSB09},
which we instantiate in this work with list-specific reasoning.
That is,
given constraints on a list whose elements belong to a theory $\mathcal T$,
we leverage our list-specific strategies to refine and encode them
into constraints in $\mathcal T$,
that we discharge to an off-the-shelf SMT solver.
In particular, while our approach resembles the modular approach
of the widely used DPLL($\mathcal T$) framework~
\cite{ganzingerDPLLTFastDecision2004},
\duelist does not interact with the DPLL algorithm directly,
nor does it delegate its backtracking logic to the underlying solver.

\begin{figure}
    \resizebox{\textwidth}{!}{\begin{tikzpicture}[node distance=15mm %
                              and 15mm %
                              ]
    \tikzstyle{basicblock} = [rectangle, align=center]
    \tikzstyle{tightblock} = [basicblock,
        text centered, minimum height=40pt, inner xsep=10pt]
    \tikzstyle{component} = [
    fill=white,
    tightblock,
    draw=black,
    ]
    \tikzstyle{hidetext} = [text opacity=0]
    \tikzstyle{overtext} = [basicblock,line width=0pt,
    text opacity=1]
    \tikzstyle{ioblock} = [basicblock, scale=0.9]
    \tikzstyle{superblock} = [dashed]
    \tikzstyle{superblockname} = [basicblock, scale=0.9]
    \tikzstyle{label} = [midway, text centered, black, scale=0.9]
    \tikzstyle{arrow} = [->, line width=0.5pt]
    \tikzstyle{innerarrow} = [arrow, draw=black!30]
    \newcommand\sectionref[1]{\centering\color{col/archi-sec-link}\footnotesize #1}

    \newcommand\superblockMargin{10pt}

    \node[ioblock] (input) {input \\ problem};

    \node[component, right=of input, xshift=-25pt] (rewrite) {
        \compRewriteBreak
    };

    \newcommand\labelLengthSolver{
        \compLengthSolver
    }

    \node[component, hidetext, right=of rewrite, xshift=20pt] (length-solver)
    {\labelLengthSolver};

    \node[right=of length-solver] (OXS->UXS) {};

    \newcommand\labelElementSolver{
        \compFixedLength
    }
    \node[component, hidetext, right=of OXS->UXS] (element-solver)
    {\labelElementSolver};

    \node[ioblock, right=of element-solver, xshift=-20pt]
        (element-solver-sat) {\SAT\\+ model};

    \node[basicblock, inner sep=6pt, below=of OXS->UXS] (smt) {SMT Solver};

    \newcommand\SMTarrow[2]{\draw[arrow, #1] (#2) -- (smt.north -| #2);}
    \newcommand\SMTcreateAnchors[1]{
        \coordinate (#1-tosmt) at ($(#1.south)+(-3pt,0)$) ;
        \coordinate (#1-fromsmt) at ($(#1.south)+(+3pt,0)$) ;
    }
    \newcommand\SMTquery[1]{
        \SMTcreateAnchors{#1}
        \SMTarrow{->}{#1-tosmt} \SMTarrow{<-}{#1-fromsmt}
    }

    \draw[arrow] (input.east) -- (rewrite.west);

    \coordinate (OXS-topleft) at
    ($(length-solver.north west) + (-\superblockMargin,\superblockMargin)$);
    \coordinate (OXS-botright) at
    ($(length-solver.east |- smt.south) + (\superblockMargin,-\superblockMargin)$);
    \coordinate (OXS-botleft) at
    (OXS-topleft |- OXS-botright);
\begin{pgfonlayer}{background}
    \draw[superblock, colOXS]
        (OXS-topleft) rectangle (OXS-botright);
    \node[superblockname, at=(OXS-botleft), anchor=north west]
        {\bfseries\color{col/OXS}{Over-approx. solver}};
\end{pgfonlayer}

    \SMTquery{length-solver}
    \node[ioblock, above=of $(length-solver -| length-solver-fromsmt)$]
        (length-solver-unsat) {\UNSAT};

    \draw[innerarrow] (length-solver-fromsmt)
    -- (length-solver-fromsmt |- length-solver.north);
    \draw[arrow] (length-solver-fromsmt |- length-solver.north)
    -- (length-solver-unsat);

    \draw[innerarrow] (length-solver-fromsmt)
    |- (length-solver.east);

    \node[overtext,at=(length-solver)] (length-solver-text) {\labelLengthSolver};

    \draw[arrow] (length-solver.east) -- (element-solver.west)
    node[label, above] {concretization}
    node[label, below] {hints};

    \coordinate (UXS-topright) at
    ($(element-solver.north east) + (\superblockMargin,\superblockMargin)$);
    \coordinate (UXS-botleft) at
        ($(element-solver.west |- smt.south) + (-\superblockMargin,-\superblockMargin)$);
\begin{pgfonlayer}{background}
    \draw[superblock, colUXS]
        (UXS-topright)
        rectangle
        (UXS-botleft);
        \node[superblockname, at=(UXS-botleft), anchor=north west]
        {\bfseries\color{col/UXS}{Under-approx. solver}};
\end{pgfonlayer}

    \tikzstyle{shiftleft} = [pos=0.4]
    \draw[arrow] (rewrite.east) -- (length-solver.west)
    node[label,above, shiftleft] {simplified}
    node[label,below, shiftleft] {constraints};

    \SMTquery{element-solver}

    \draw[innerarrow] (element-solver-fromsmt) |- (element-solver.east);
    \draw[arrow] (element-solver.east) -- (element-solver-sat);

    \draw[innerarrow] (element-solver-fromsmt)
    -- (element-solver-fromsmt |- element-solver.north);

    \draw[arrow] (element-solver-fromsmt |- element-solver.north)
    -- ++(up:2*\superblockMargin)
    -| (length-solver.40)
    node[label, pos=0.25, above]{feedback}
    ;

    \node[overtext,at=(element-solver)] (element-solver-text)
    {\labelElementSolver};

    \draw[component, fill opacity=0] (length-solver.west |- smt.south)
    rectangle (smt.north -| element-solver.east);

\end{tikzpicture}}
    \caption{Overview of \duelist's architecture}
    \Description{Schema described in the corresponding section}
    \label{fig:architecture}
\end{figure}
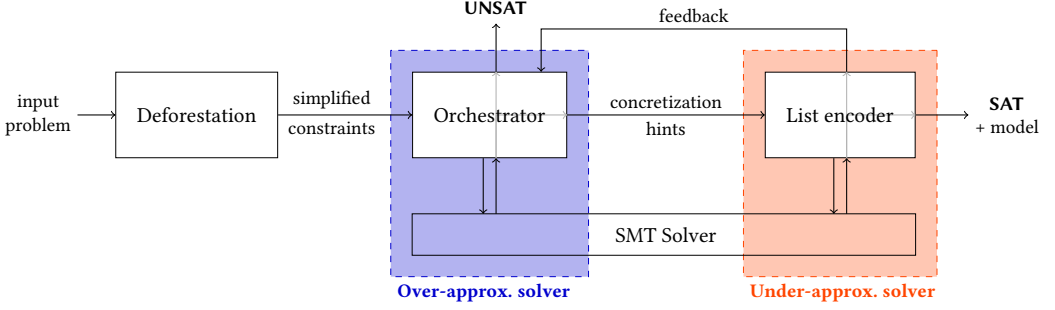

We present in \Cref{fig:architecture} an overview of our approach, and describe below the main components.

\paragraph{Input problem}
The input of \duelist is a sequence of constraints
encoded in the \duelist language described in \Cref{sec:language}.
These constraints can relate to lists, e.g., using \len, or the \map and \fold
combinators, but also to the other theories supported by the underlying
solver.

\paragraph{Deforestation}
Before entering its main loop,
\duelist first applies list-specific rewriting passes (\Cref{sec:optimizations:rewrite}) to simplify the considered constraints.
This pass removes intermediate lists, akin to~\citet{DBLP:journals/tcs/Wadler90}'s deforestation used in the compilation of functional languages.

\paragraph{\OXS}
The \oxs relies on list-specific abstractions to encode an over-approximation of the input problem in theories natively supported by the SMT solver.
These abstractions relate to the shape of the lists considered, for instance, to reason about the image of a fold (\Cref{sec:optimizations:reachability})
or to abstract lists through their length, enabling efficient arithmetic reasoning (\Cref{sec:length:ox}).
These abstractions take inspiration from previous works in static analysis of arrays and strings~\cite{blanchet2002design, kastner:inria-00528600,gopan2004numeric,DBLP:conf/pldi/VenetB04,moraZ3str4MultiarmedString2021,DBLP:conf/sas/JournaultMO18}.

Importantly, the abstractions considered by the \oxs must be \emph{sound}, that is,
they must generate over-approximations of the input problem. Under this condition,
we are therefore able to conclude about the unsatisfiability of the input constraints when the underlying SMT solver can establish the unsatisfiability of the abstracted problem.
This soundness condition can also be expressed by contraposition, stating that if the
input problem is satisfiable, then the over-approximating abstraction must also be
satisfiable. This alternate formulation can be more convenient to use when designing
abstractions.

In the case where the SMT solver does not establish unsatisfiability,
either the input problem is satisfiable, or the abstraction is too coarse to determine unsatisfiability.
The \oxs then leverages its abstractions to pass concretization hints
to the second component, the \uxs.

\paragraph{\UXS}
The \uxs simplifies the input problem by only reasoning about a subset of the search space.
To do so, it relies on the concretization hints passed by the \oxs to translate
the problem to constraints natively supported by the underlying solver, which must be equivalent
to the original constraints for the input space subset considered.
This process heavily depends on the concretization hints, which can specify, e.g., to only consider
lists of a given length $n$, allowing to replace a symbolic list by $n$ symbolic variables (\Cref{sec:length:ux}), or to
force a list shape to erase specific fold operations (\Cref{sec:fold:neutral}).

If the solver is able to determine the satisfiability of the constraints on the restricted input space,
then we can conclude about the satisfiability of the input constraints. In this case, the list encoder
will translate the model returned by the SMT solver to directly relate to the symbolic variables
defined in the input problem. This includes, e.g., reconstructing an idiomatic list when the list
was encoded as $n$ symbolic variables.

\paragraph{Feedback loop.}
When the \uxs is unable to establish the satisfiability of the problem, it provides feedback
to the \oxs, allowing it to \emph{refine} its abstractions and therefore reason about a finer-grained over-approximation of the problem.
The orchestrator implements the abstraction-refinement loop.
It repeats the process until it is able to conclude about the (un)satisfiability of the input problem, or a timeout is reached.

\paragraph{Notations}
We denote concrete values in bold (\eg, \conc 0, \conc{[1,2,3]})
to distinguish them from symbolic values and variables.

\duelist answers queries with \SAT, \UNSAT, or \UNKNOWN,
while the underlying SMT solver answers \SMTSAT, \SMTUNSAT, or \SMTUNKNOWN.
Models provided by solvers are denoted by the letter $\model{M}$,
or described explicitly as a mapping from variables to values,
such as $l \mapsto \conc{[1,2]}$.
We access the concrete value corresponding to a variable $x$
in a model $\model{M}$ by writing $\modelAccess{\model M}{x}$.

Given a symbolic list \smtlib{l}, we write $\lenvar{\smtlib l}$ for
the symbolic variable representing its length.
If the length of a list \smtlib{l} is statically known to be a constant
$\conc n$, we represent its elements as variables $\liln{\smtlib l}n$.

\subsection{\duelist by Example}
\label{sec:architecture:example}

To provide the reader with a high-level intuition of our approach,
we now apply it step-by-step to the running example from \Cref{fig:running.ml}.
For the sake of presentation, this example focuses on the interaction loop
between the orchestrator and the list encoder, and does not apply any heuristics
or optimizations.
We summarize the main steps that \duelist takes
in \Cref{fig:main-loop}, and we give more details below.

\begin{figure}
    \resizebox{\textwidth}{!}{\begin{tikzpicture}[node distance=25mm %
                              and 15mm %
                              ]
    \tikzstyle{basicblock} = [rectangle, align=center]
    \tikzstyle{titleblock} = [basicblock, scale=0.9]
    \tikzstyle{xblock} = [basicblock, draw]
    \tikzstyle{oxblock} = [xblock,text=col/OXS,draw=col/OXS]
    \tikzstyle{uxblock} = [xblock,text=col/UXS,draw=col/UXS]
    \newcommand\leftColWidth{180pt}
    \newcommand\rightColWidth{100pt}

    \tikzstyle{label} = [midway, text centered]
    \tikzstyle{label/query} = [label, scale=0.9]
    \tikzstyle{label/comm} = [label, scale=0.8, right]
    \tikzstyle{arrow} = [->, line width=0.5]

    \newcommand\titleabove[3][]{
        \node[titleblock, at=(#2.north west), anchor=south west, #1] {#3};
    }
    \newcommand\OXSabove[1]{\titleabove[text=col/OXS, align=left]{step#1}
        {Step #1:\\Over-approx. solver}}
    \newcommand\UXSabove[1]{\titleabove[text=col/UXS, align=left]{step#1}
        {Step #1:\\Under-approx. solver}}

    \node[basicblock,draw,anchor=west] (input) {
        \begin{minipage}{\leftColWidth+\rightColWidth}
\begin{OCaml}
    let custody_score = List.fold (+) 0 (List.map f children) in
    custody_score > 2 && List.len children > 0
\end{OCaml}
        \end{minipage}
    };
    \titleabove{input}{Step 1: Input Problem}

    \node[basicblock, right=of input] (smt) {\phantom{M}};

    \newcommand\SMTcreateAnchors[1]{
        \coordinate (#1-tosmt) at ($(#1.east)+(0,+3pt)$) ;
        \coordinate (#1-fromsmt) at ($(#1.east)+(0,-3pt)$) ;
    }
    \newenvironment{tightalign}{
        \begin{minipage}{\rightColWidth}
            \setlength{\jot}{-1pt}
            }{
        \end{minipage}
    }
    \newcommand\SMTarrowTo[2]{
        \draw[arrow, ->] (#1-tosmt) -- (smt.west |- #1-tosmt)
        node[label/query,above] {
            \begin{tightalign}
                #2
            \end{tightalign}
        };
    }
    \newcommand\SMTarrowFrom[3][]{
        \draw[arrow, <-] (#2-fromsmt) -- (smt.west |- #2-fromsmt)
        node[label/query,below,#1] {#3};
    }

    \setlist[itemize]{leftmargin=*, label=\textbullet}

    \node[oxblock, below=of input.west, anchor=west] (step2) {
        \begin{minipage}{\leftColWidth}\begin{itemize}
            \item encode length constraint:\\$\smtlib{custody_score} > 2 \land \lenvar{\smtlib{children}} > 0$
            \item call SMT
        \end{itemize}\end{minipage}
    };
    \OXSabove{2}

    \draw[arrow] (input.south -| step2.north) -- (step2.north);

    \SMTcreateAnchors{step2}
    \SMTarrowTo{step2}{
        \begin{align*}
            &\smtlib{custody_score} > 2\\
            &\lenvar{\smtlib{children}}> 0
        \end{align*}
    }
    \SMTarrowFrom[yshift=8pt]{step2}{\begin{tightalign}\begin{align*}
            \SMTSAT:{}&\lenvar{\smtlib{children}}\mapsto\conc 1, \\
                      &\smtlib{custody_score}\mapsto\conc{10} \\
        \end{align*}\end{tightalign}
    }

    \node[uxblock, below=of step2.west, anchor=west] (step3) {
        \begin{minipage}{\leftColWidth}\begin{itemize}
            \item encode $\ml{children}$ as
                $[\li{\smtlib{children}}1]$
            \item call SMT with refined query
        \end{itemize}\end{minipage}
    };
    \UXSabove{3}

    \draw[arrow] (step2.south) -- (step3.north)
    node[label/comm] {length candidate \#1};

    \SMTcreateAnchors{step3}
    \SMTarrowTo{step3}{
        \begin{align*}
            &\li{\smtlib{children}'}1 = \app f{\li{\smtlib{children}}1} \\
            &0 + \li{\smtlib{children}'}1 > 2\\
            &\conc 1 > 0
        \end{align*}
    }
    \SMTarrowFrom{step3}{\SMTUNSAT}

    \node[oxblock, below=of step3.west, anchor=west] (step4) {
        \begin{minipage}{\leftColWidth}\begin{itemize}
            \item $\lenvar{\smtlib{children}}\mapsto \conc 1$
                does not work
            \item call SMT
        \end{itemize}\end{minipage}
    };
    \OXSabove{4}

    \draw[arrow] (step3.south) -- (step4.north)
    node[label/comm] {feedback};

    \SMTcreateAnchors{step4}
    \SMTarrowTo{step4}{
        \begin{align*}
            &\smtlib{custody_score} > 2 \\
            &\lenvar{\smtlib{children}} > 0 \\
            &\lnot\left(\lenvar{\smtlib{children}} = 1\right)
        \end{align*}
    }
    \SMTarrowFrom{step4}{$\SMTSAT: \lenvar{\smtlib{children}}\mapsto\conc 2, \ldots$}

    \node[uxblock, below=of step4.west, anchor=west] (step5) {
        \begin{minipage}{\leftColWidth}\begin{itemize}
            \item encode \ml{children} as
                $[\li{\smtlib{children}}1, \li{\smtlib{children}}2]$
            \item call SMT
        \end{itemize}\end{minipage}
    };
    \UXSabove{5}

    \draw[arrow] (step4.south) -- (step5.north)
    node[label/comm] {length candidate \#2};

    \SMTcreateAnchors{step5}
    \SMTarrowTo{step5}{
        \begin{align*}
            &\li{\smtlib{children}'}1 = \app f{\li{\smtlib{children}}1} \\
            &\li{\smtlib{children}'}2 = \app f{\li{\smtlib{children}}2} \\
            &0 + \li{\smtlib{children}'}1 + \li{\smtlib{children}'}2 > 2\\
            &\conc 2 > 0
        \end{align*}
    }
    \SMTarrowFrom[yshift=8pt]{step5}{\begin{tightalign}\begin{align*}
            \SMTSAT:{}&\li{\smtlib{children}}1\mapsto\conc{Sole}, \\
                    &\li{\smtlib{children}}2\mapsto\conc{Shared} \\
        \end{align*}\end{tightalign}
    }

    \node[below of=step5, yshift=20pt] (output) {
        $\SAT: \ml{children} \mapsto \conc{[Sole, Shared]}$
    };
    \draw[arrow] (step5.south) -- (output.north);

    \coordinate (smt-topleft) at (smt.west |- step2.north);
    \coordinate (smt-botright) at  (smt.east |- step5.south);
    \draw (smt-topleft) rectangle (smt-botright);

    \node at ($(smt-topleft)!0.5!(smt-botright)$) [inner sep=0pt]
    {\rotatebox{90}{SMT solver}};

\end{tikzpicture}}
    \caption{Example of abstraction-refinement loop performed by \duelist}
    \Description{Schema described in the corresponding section}
    \label{fig:main-loop}
\end{figure}

We simplify the running example by setting respectively \ml{custody_score_threshold} = \conc 2 and \ml{child_count_threshold} = \conc 0.
In practice, \duelist is able to reason about this example when these two variables
are symbolic. This problem is passed as an input to \duelist (Step~1);
our solver will try to determine whether it is satisfiable, and if so, return a model for
the list \ml{children}.

\duelist first passes the problem to the \oxs (Step~2), which abstracts over the list
elements of the constraints to generate an SMT-compatible query. In this example,
we rely on a length abstraction: we replace occurrences of the list \ml{children}
by $\lenvar{\smtlib{children}}$, and adapt the constraints accordingly. For instance,
the expression \ml{List.len children} simply becomes $\lenvar{\smtlib{children}}$,
while the length abstraction is too coarse to extract information about the
result of a \ml{fold} operation; we replace its result by a new, unconstrained
symbolic variable, \ml{custody_score}.

Leveraging this abstraction, \duelist is then able to encode the problem
as the SMT query $\smtlib{custody_score} > 2 \wedge \lenvar{\smtlib{children}} > 0$.
The SMT solver determines that this query is satisfiable, and returns a model where
$\lenvar{\smtlib{children}} \mapsto \conc 1$ and $\smtlib{custody_score} \mapsto \conc{10}$.
From this model, the \oxs extracts and passes to the \uxs the concretization hint
$\lenvar{\smtlib{children}} \mapsto \conc 1$, instructing it to restrict the search space
to cases where the list \ml{children} only contains one element.

Relying on this hint, the \uxs (Step~3) then partially concretizes the symbolic list \ml{children},
replacing it by the list $[\li{\smtlib{children}}1]$, where $\li{\smtlib{children}}1$ is a fresh
symbolic variable. It then propagates this change to the different list operators occurring
in the problem: the expression \ml{List.map f children} is replaced by the application
of the function \ml{f} to $\li{\smtlib{children}}1$, creating a fresh, intermediate variable
$\li{\ml{children'}}1$. The \ml{fold} application is unfolded, while \ml{List.len children}
becomes the constant $\conc 1$. The resulting query is then passed to the SMT solver, which
returns \SMTUNSAT: indeed, in this example, the function \ml{f} passed as an argument to \ml{map}
returns $\conc 0, \conc 1$, or
$\conc 2$. Therefore, the constraint $\smtlib{custody_score} > 2$, which became
$0 + \li{\smtlib{children}'}1 > 2$, is unsatisfiable. As the \uxs cannot conclude about
the satisfiability of the problem, it gives control back to the \oxs while informing it
that the list \ml{children} cannot be of length $\conc 1$.

Leveraging feedback from the \uxs, the \oxs (Step~4) can then \emph{refine} its abstraction by
explicitly excluding cases where $\lenvar{\smtlib{children}} = \conc 1$; it does so by adding the
negation of this constraint to its SMT query before repeating the process. The SMT solver
returns \SMTSAT with a model where $\lenvar{\smtlib{children}} \mapsto \conc 2$, which the
\oxs transfers to the \uxs. Restricting its search to cases where
$\lenvar{\smtlib{children}} = \conc 2$, the \uxs (Step~5) is then able to determine the
satisfiability of the problem. It finally lifts the model returned by the SMT solver
to return a model pertaining to list variables, where
$\ml{children} \mapsto \conc{[Sole, Shared]}$.

\section{Reasoning through Length Abstractions}
\label{sec:length}

Length abstractions are a common component in many program analysis tools,
abstracting over, e.g., string or array values to only reason about their lengths~\cite{DBLP:conf/pldi/VenetB04,moraZ3str4MultiarmedString2021,DBLP:conf/sas/JournaultMO18}.
In this section, we show how to integrate length abstractions in our framework, first
to define sound list over-approximations (\Cref{sec:length:ox}), then to restrict
the input space (\Cref{sec:length:ux}) before finally describing how we use feedback to refine
the abstraction (\Cref{sec:length:feedback}).

\subsection{Over-approximation by Length Abstraction}
\label{sec:length:ox}

We rely on length abstractions to erase list expressions, creating an encoding
into the boolean, integer, and datatypes theories of SMT-LIB.
This encoding enables querying the underlying SMT solver to reason about list
length models.
We formally present rules for our translation procedure in
\Cref{fig:extraction-rules}.

The translation operates in a top-down manner, by recursing into subterms.
The core operation is \Rule{List-Var}, which replaces a list variable $\sym l$
by a symbolic integer $\lenvar l$ representing its length.
Non-list variables (\Rule{Var}) and literals (\Rule{Lit}) are left unchanged by the translation.

\Rule{List-Expr} replaces lists of statically known length \conc n, created through the \smtlib{List.value} operator, with their length \conc n.

\Rule{BinOp} translates binary operators, such as integer addition or boolean comparison.
To do so, it recursively translates both operands, and returns the application of the
operator to the translated terms.
Our translation scheme also supports datatypes and corresponding operations, that are omitted here for brevity.

\begin{figure}
    \centering
    \begin{mathpar}
      \inferrule[List-Var]{{\sym l} \text{ is a list variable}}{\extractLen{\sym l}{\lenvar{l}}}

      \inferrule[List-Expr]{}{\extractLen{[\liln en]}{\conc n}}

      \\

      \inferrule[Var]{{\sym v} \text{ is a non-list variable}}{\extractLen{\sym v}{\sym v}}

      \inferrule[Lit]{\conc k \text{ is a non-list literal}}{\extractLen{\conc k}{\conc k}}

      \inferrule[BinOp]{\extractLen{e_1}{l_1}\quad\extractLen{e_2}{l_2}}{\extractLen{e_1 \bowtie e_2}{l_1 \bowtie l_2}}

      \\

      \inferrule[Len]{\extractLen{e}{l}}{\extractLen{\applen{e}}{l}}

      \inferrule[Map]{\extractLen{e}{l}}
                  {\extractLen{\appmap fe}{l}}

      \inferrule[Fold]{}{\extractLen{\appfold fie}{{\sym sym}_{\appfold fie}}}
    \end{mathpar}
    \Description{Length abstraction translation rules} %
    \caption{Length abstraction translation rules}
    \label{fig:extraction-rules}
\end{figure}

\Rule{Len} translates occurrences of the \len combinator. It applies to a list expression,
which is recursively translated, thus yielding its length abstraction. The \len combinator
can therefore directly return the result of this translation.

The \map combinator takes a function $f$ and a list expression $l$, and applies $f$
to all elements of $l$. It therefore does not modify the length of $l$, and can
directly return the length abstraction recursively computed for $l$, as captured
by \Rule{Map}.

\Rule{Fold} contains several subtleties. As, in the general case, we cannot infer
information about the result of a \fold from only the length abstraction of its
list argument, our translation scheme returns an unconstrained symbolic variable
whose type corresponds to the return type of the \fold.
Returning a fresh symbolic variable for each \fold occurrence would however be too
coarse, and hamper precise reasoning. Consider for instance the constraint
$\appfold fie = \appfold fie$. If we returned fresh variables at each \Rule{Fold}
application, this constraint would be translated to $sym_1 = sym_2$, losing the
relationship between the two terms and failing to detect that this equality always
holds. We therefore associate a symbolic variable to each \fold expression, which
we denote as $sym_{\appfold fie}$.

Crucially, the correctness of the \fold translation relies on two technical details.
First, the SMT-LIB language does not support variable redeclaration, which prevents
shadowing. This therefore ensures that two syntactically equal terms are indeed
semantically equivalent. Second, the \Rule{Fold} rule does not recursively
translate subterms before associating the fold expression to a symbolic variable,
but rather operates directly on the terms in the input problem.
Consider for instance terms $\appfold fil$ and $\appfold fi{(\appmap gl)}$, where
$l$ is a list variable. Translating the \fold list arguments would
lead to the same term $\lenvar l$ on both sides, and thus to the same symbolic
variable when translating the entire \fold expression, incorrectly capturing that both terms are always equal.

\paragraph{Well-Typedness}
To capture the validity of the translation, we need to show that the translated
problem will be accepted by the SMT solver, and therefore that the resulting
program is \emph{well-typed}. To do so, we rely on the following lemma,
which states that any list expression has indeed been replaced by
an integer expression, which in practice corresponds to the length abstraction.
The typing rules are straightforward, and omitted for brevity.

\begin{lemma}
  \label[lemma]{lemma:type}
  For any expressions $e, e'$, any typing context $\Gamma$, and $e \overset{*}{\to} e'$
  \begin{enumerate}
  \item If $\Gamma \vdash e : list~t$, then $\Gamma \vdash e' : int$
  \item If $\Gamma \vdash e : t$ and $t$ is not a list, then $\Gamma \vdash e' : t$
  \end{enumerate}
\end{lemma}

The proof is performed by induction on the translation derivation. The intuition is that
lists are either of constant length, and translated to $int$ literals,
or they are abstract, so one can only interact with them through the \map, \fold,
or \len combinators, and the \fold combinator cannot return a list.

\paragraph{Soundness}
To conclude about the unsatisfiability of the input problem using length abstractions, we rely on
the following lemma stating that the translation correctly over-approximates it.

\begin{lemma}
  \label[lemma]{lemma:soundness}
  For any expressions $e, e'$ such that $e \overset{*}{\to} e'$,
  if $e'$ is \SMTUNSAT, then $e$ is \UNSAT.
\end{lemma}

To establish this property, we prove the contraposition: if $e$ is \SAT, then $e'$ is \SMTSAT.
The proof proceeds again by induction on the translation derivation. The core part of the proof
relates to \Rule{List-Var}: if $e$ is \SAT, then there exists a list literal that models the symbolic list $l$
and we can apply \Rule{List-Expr} to it to build an integer model for the symbolic variable $\lenvar l$.

\paragraph{Interacting with the SMT Solver}

\Cref{lemma:type} also guarantees that list expressions in the input problem have been
rewritten, which allows us to query the SMT solver with the translated problem.
If the SMT solver returns \SMTUNSAT, then we can directly conclude from \Cref{lemma:soundness}
that the input problem is \UNSAT, and return this result to the user.
Otherwise, if the SMT solver returns \SMTSAT, it also provides a model with concrete
values for all symbolic variables. In this case, we extract values corresponding
to the length abstraction of each list variable in the context, and pass them
to \duelist's second component, the \uxs.

\subsection{Fixed-Length Under-approximations}
\label{sec:length:ux}

The \uxs aims to leverage restrictions to the search space to encode the
input problem as an SMT query, removing list-related terms and constraints
unsupported by the solver. When considering length abstractions, a natural
under-approximation is to only consider fixed-length lists.
We describe this encoding formally in \Cref{fig:conc-rules}
and detail its rules below.

\begin{figure}
    \centering
    \begin{mathpar}
      \inferrule[Conc-List]{{\sym l} \text{ is a list variable} \quad \lenvar l \mapsto \conc n}{\rewriteFixed{\sym l}{[\liln ln]}}

      \inferrule[Conc-Len]{\rewriteFixed{e}{[\liln en]}}{\rewriteFixed{\applen{e}}{\conc n}}

      \inferrule[Conc-Map]{\rewriteFixed{e}{[\liln en]}}
      {\rewriteFixed{\appmap fe}{[f\; e_{\conc 1}, \ldots, f\; e_{\conc n}]}}

      \inferrule[Conc-Fold]{\rewriteFixed{e}{[\liln en]}}
      {\rewriteFixed{\appfold fie}{f~(~\ldots~(f~(f~i~e_{\conc 1})~e_{\conc 2})~\ldots)~e_{\conc n}}}
    \end{mathpar}
    \Description{Selected rules for fixed-length under-approximation} %
    \caption{Selected rules for fixed-length under-approximation}
    \label{fig:conc-rules}
\end{figure}

For each symbolic list $l$ in the input problem, the \uxs receives a concretization
hint that specifies an admissible length $\conc n$. Based on this information,
the list $l$ is transformed into \conc n symbolic variables $\liln ln$ (\Rule{Conc-List}).
Similarly to the translation scheme presented in the previous section,
\duelist then recursively rewrites all terms in the problem to replace
all occurrences of list combinators, as formally described in \Cref{fig:conc-rules}.

As the length of each symbolic list is fixed by the concretization hints
obtained from the \oxs, each list expression has a statically known length during the
translation. Thus, each \len combinator can be directly replaced by the corresponding
constant (\Rule{Conc-Len}).

A \map combinator applies a function $f$ to all elements of a list expression $e$.
\duelist first translates the argument $e$, then creates a new list of the same
length, where $f$ is applied pointwise (\Rule{Conc-Map}).

As the length of all list expressions is statically known, translating \fold can
be done in a straightforward manner, by \emph{unfolding} the operation. Thus,
an expression $\appfold{f}{i}{e}$, where $e$ is translated to $\liln en$, can
be translated as
$\apptwo f
    {(\ldots \apptwo f{(\apptwo f{\sym{i}}{\li e1})}{\li e2} \ldots)}
    {\li en}$ (\Rule{Conc-Fold}).

\paragraph{Interacting with the SMT Solver}

The translation presented above removes occurrences of list combinators, however,
it still preserves list structures, even though the lists are of fixed length.
In practice, \duelist keeps information about the length and variables of each
list sub-expression internally, and relies instead on auxiliary symbolic variables
to entirely remove list values. For example, consider the translation of a $\appmap fe$
expression previously presented. Instead of translating $e$ to a value $[\liln en]$,
\duelist instead tracks the symbolic expressions $\liln en$ corresponding to the
result of the translation. It then generates $n$ fresh symbolic variables $\liln {l'}n$,
alongside the constraints $\li{l'}{i} = f\; \li{e}{i}$ in the translated SMT query.
Finally, it internally keeps track that translating $\appmap fe$ corresponds
to the list $[\liln {l'}n]$, enabling the rest of the translation to proceed.

Having erased all notions of lists, the resulting query can then be passed to
standard SMT solvers. If the solver returns \SMTSAT,
then a solution to the initial problem has been found and \duelist returns \SAT.
It also lifts the SMT model $\modelSMT$ to provide a model for the input problem,
which included list variables. Leveraging its tracking of the relationship between
each list variable $l$ and the symbolic variables $\liln ln$,
it associates to $l$ the model $[\modelAccess\modelSMT{\li l1}, \ldots, \modelAccess\modelSMT{\li ln}]$, and removes all auxiliary variables generated as part of the translation.
Other symbolic variables are kept as-is from $\modelSMT$, and directly propagated to the user.

\subsection{Feedback Loop}
\label{sec:length:feedback}
When the SMT solver returns \SMTUNSAT in the \uxs, \duelist concludes that no solution
to the input problem exists with the fixed lengths passed as concretization hints.
This feedback allows \duelist to refine the length abstraction, by excluding these lengths
from the SMT query generated by the \oxs.

Consider for instance the problem
$\applen {\sym l} < 10 \land \appfold{(+)}{0}{\sym l} = 1$.
As described in \Cref{sec:length:ox}, the length abstraction approach
would translate it to $\lenvar l < 10 \land sym_{\appfold{(+)}{0}{\sym l}} = 1$,
which is satisfiable and could return a model where $\lenvar l \mapsto 0$.
Passing this hint to the \uxs would yield an \SMTUNSAT result, and feedback
that the problem is not satisfiable when $l$ is an empty list. The \oxs
would then add the constraint $\lnot(\lenvar l = 0)$ to its previously translated
query, generate and pass a new length model to the \uxs, which would allow \duelist
to conclude about the satisfiability of the problem.

In the general case, this feedback loop would continue, iteratively refining the abstraction
until either satisfiability or unsatisfiability of the problem can be established, or until a timeout is reached. In the latter case, \duelist would return \UNKNOWN.

Note that the length exploration strategy leaves free rein to the SMT solver, which may (and in practice, does) explore lengths in non-increasing order.
An alternative strategy is to force the solver to search for increasing lengths.
While this strategy guarantees to find the shortest lists in satisfiable instances, it may not fully exploit inferred length information.
Binary search strategies are left as future work, as they would require a semantic analysis of monotonicity to be applicable.

\section{Abstracting over Fold Combinators}
\label{sec:fold}

While length abstractions are an efficient method to reason about lists or arrays, they do not
allow reasoning about the \emph{contents} of a list. To circumvent this issue, several program
analysis tools complement them with specific abstractions such as array
smashing~\cite{blanchet2002design, kastner:inria-00528600} (also known as array
summarization~\cite{gopan2004numeric}), where a single symbolic variable represents all values
in an array.

Smashing abstractions particularly shine on arrays, where most operations consist in accessing or updating
array elements. They are however less efficient in our setting, where a coarse abstraction for
list elements can lead to highly imprecise abstractions for \fold combinators.
Drawing inspiration from such techniques, we present in this section fold-specific abstractions which
aim to retain a high level of precision while enabling abstract reasoning about lists.

\subsection{List Summarization}
\label{sec:fold:neutral}

We first show how to leverage the algebraic structure of the \fold combinator to propose
a variant of array summarization applicable to lists. To describe our approach
on an example, consider the following problem:

\[
  \applen l > 100 \wedge
  \appfold{(\applambdatwo{acc}{x}{(acc + x)})}{0}{l} > 50
\]

The \fold expression computes the sum of all elements in the list $l$.
This query is therefore trivially satisfiable: for instance, any list of length greater than 100
that contains only ones is a solution.

Unfortunately, given the length requirement on the list $l$, the approach
presented in \Cref{sec:length} would be inefficient when applied to this problem.
Indeed, the length abstraction used in the \oxs would instruct the \uxs to restrict
its search to cases where $l$ has a large fixed size. In turn, this would
create a large number of intermediate variables and constraints to handle all
elements of $l$, hampering scalability for more complex problems.

To tackle this issue, we investigate conditional rewritings that allow us to
better encode the problem as an SMT query. Notice that, due to the associativity
and commutativity of the addition, for any $\liln ln$, the expression
 $\appfold {(+)}0{[\liln ln]}$ is equivalently computed as $\appfold {(+)}0{[(\li l1 + \ldots + \li ln), 0, \ldots, 0]}$, (with $\conc{n-1}$ zeros).
In our example problem, since there are no other constraints on $l$ than the \fold and its length,
the input problem is satisfied by some list $\conc{[\liln ln]}$, if and only if it is also satisfied by $\conc{[\li l1 + \ldots + \li ln, 0, \ldots, 0]}$.

Thus, we can abstract $l$ as two symbolic integers: its length $\lenvar l$ and its head $h$.
If the SMT solver returns a model $\lenvar l \mapsto \conc n, h \mapsto \conc h$,
we can rebuild a list model starting with the head $\conc h$ and padded with the neutral element $\conc 0$ to reach the required length.
In our setting, abstracting $l$ as the symbolic variables $\lenvar l$ and $h$
yields the simplified constraints $\lenvar l > 100 \land h > 50$, which SMT solvers easily determine
to be \SMTSAT, allowing \duelist to conclude about the satisfiability of the problem.

\paragraph{Reducing \fold{}s on functions with fixpoints}%
The \fold constraint in the example above can be simplified because of two properties.
First, any resulting value can be reached by the \fold's function using the initial value \smtlib{0} and the head element of the list.
Second, the list can be padded by a neutral element for the \fold's function
(here, the value $\conc 0$) so that once the resulting value is reached after one \fold iteration,
summing the rest of the list does not change it.
\newcommand\hd{h\xspace}

We can generalize this representation of the sum to any function that respects these two properties.
Let $f : \beta \to \alpha \to \beta$ be the functional used in the \fold, and $i: \beta$ be the initial element of the \fold.
Assuming that \Cref{eqn:neutral} holds, any list $l: \listtype\alpha$ can be transformed into $l' = [\hd, e, \ldots, e]$, where $l$ and $l'$ have the same length.%

\begin{equation}
    \label{eqn:neutral}
    \exists e: \alpha, \forall B: \beta, \exists \hd: \alpha,
    \apptwo fi\hd = B \land \apptwo fBe = B.
\end{equation}

Additionally, the \fold operator can then be rewritten in a length-independent expression thanks to \Cref{eqn:reduction}.
Note that we need to distinguish between the case where the list is empty, in which case the \fold operator directly
returns the initial value $i$, and the case where the list is non-empty and the function $f$ applies.

\begin{equation}
    \label{eqn:reduction}
    \appfold fil = \appfold fi{[\hd, e, \ldots, e]} = \appite{(\applen l = 0)}{i}{(\apptwo f i h)}
\end{equation}

To summarize, if \Cref{eqn:neutral} applies, any list can be reduced to a list
of a specific shape, characterized by the head element and its length.
The $\fold f$ is equivalent to a length-independent expression,
which avoids the need for a large number of symbolic variables and constraints
when the length model instructs the \uxs to consider large lists, therefore
greatly improving the performance of the solver.

\paragraph{Implementation in \duelist}
When an expression $\appfold f i l$ occurs in the input problem,
the \oxs instantiates \Cref{eqn:neutral} with the sub-expressions $f$ and $i$,
and queries the SMT solver.
If it answers \SMTUNSAT, then the heuristic cannot be used.
If it answers \SMTSAT, the \oxs retrieves a model for the neutral element $e$,
and passes it to the \uxs, along with the information that the \fold expression
can be abstracted through symbolic variables for the list's length and head.

Using this hint, the \uxs then creates fresh symbolic variables for \sym h and \lenvar l,
and rewrites the \fold expression according to \Cref{eqn:reduction} before querying the SMT solver.
If it answers \SMTUNSAT, then the input problem is unsatisfiable,
and \duelist can answer \UNSAT.
If it answers \SMTSAT, with a model $\modelSMT$ for $h$ and $\lenvar l$,
the list encoder can leverage $\modelSMT$ to reconstruct a list model.
If $\modelAccess\modelSMT{\lenvar l} = \conc 0$,
then $l \mapsto \conc{[]}$ satisfies the constraints.
If $\modelAccess\modelSMT{\lenvar l}$ is some nonzero integer \conc n,
then
$l \mapsto \big[
    \modelAccess\modelSMT{h},
    \underbrace{\conc e, \ldots, \conc e}_{\text{\conc{n-1} times}}
\big]$
satisfies the constraints.

\subsection{\smtlib{fold} Reachability}
\label{sec:optimizations:reachability}

To efficiently reason about the unsatisfiability of input problems,
we now propose another heuristic that reasons on the image of the
function $f$ passed as a \fold argument.

Consider the following example,
where a \fold retrieves the head of a non-empty list
\[
    \begin{gathered}
        \applen l > 0 \\
        \smtlib{None} = \appfold
        {(\applambdatwo{acc}{x}{\appmatch{acc}
        \;|\smtlib{None} \to \smtlib{Some x}
        \;|\smtlib{Some _} \to acc})}
        {\smtlib{None}}l
    \end{gathered}
\]
These constraints are unsatisfiable
because for any non-empty list $l$,
the \fold combinator will return a \smtlib{Some} expression.
\duelist can see this automatically
without creating a length model
and without unrolling the \fold:
under the assumption that the list is non-empty, and thus that the folded function $f$
is applied at least once, it suffices to check that
\smtlib{None} does not belong to the image of $f$.

Let us denote as $\phi$ the input problem, and as
$\phi[x\backslash y]$ the input problem where expression $x$
has been replaced by expression $y$.
When encountering a $\appfold fil$ expression, we can over-approximate
it as the image of $f$ when $l$ is non-empty, and rewrite it as the expression $i$
when $l$ is empty.

Formally, we thus rewrite the input problem $\phi$ to consider the following
alternative problem:
\begin{equation}
    \label{eqn:reachability:necessary}
    \Big(\applen l = 0 \land \varphi\big[(\appfold fil)\backslash i\big]\Big)
    \lor \Big(\applen l \neq 0 \land \exists~x~y, \varphi\big[(\appfold fil)\backslash (\apptwo fxy)\big]\Big)
\end{equation}

\duelist provides this heuristic as part of its \oxs. When encountering a \fold
expression, it first rewrites the problem into its alternate formulation described
in \Cref{eqn:reachability:necessary}. It then relies on the length abstraction
presented in \Cref{sec:length:ox} to abstract over the newly added $\applen l$
expressions, and therefore enable encoding the problem as an SMT query.

As this rewriting abstracts over all possible values for the \fold expression,
if the alternative problem is unsatisfiable, then so is the initial problem $\phi$.
If the SMT solver answers \SMTUNSAT on the rewritten problem, \duelist can thus safely conclude
that the initial problem is \UNSAT.
Otherwise, when the SMT solver returns \SMTSAT, the abstraction is too coarse to provide
meaningful hints to the \uxs, and the \oxs continues its solving procedure with different
heuristics.

\section{Simplifying and Optimizing SMT Queries}
\label{sec:optimizations}

To reason more efficiently about input problems, SMT solvers commonly
provide a variety of theory-specific simplification rules and
general optimizations. In this section, drawing inspiration from a compilation
technique for functional languages, we first propose simplifications
specific to lists (\Cref{sec:optimizations:rewrite}).
We then discuss the applicability of common techniques to improve the performance of the SMT solver (\Cref{sec:optimizations:technical}).

\subsection{SMT Variable Reduction Through Deforestation}
\label{sec:optimizations:rewrite}

\emph{Deforestation}, initially proposed by~\citet{DBLP:journals/tcs/Wadler90}, is a program
transformation which removes computations of intermediate algebraic datatypes.
While this optimization is originally targeting compilation of functional languages, it is also beneficial in \duelist's context: deforestation enables a significant reduction of the
variables passed to the SMT solver.
Intuitively, this optimization leverages the compositionality of combinators to
simplify list constraints.

\begin{figure}
  \begin{minipage}{.5\textwidth}
\begin{lstlisting}[
    style=ocaml,
    firstline=2,
    lastline=8,
    ]
  List.fold (+) 0
    (List.map (fun child ->
         match child.custody with
         | Sole -> 2
         | Shared -> 1
         | None -> 0)
        children)
\end{lstlisting}
  \Description{Source code} %
    \caption{Running example}
    \label{fig:deforestation:og_prog}
  \end{minipage}\begin{minipage}{.5\textwidth}
\begin{OCaml}
    List.fold (fun acc child ->
        acc +
        (match child.custody with
            | Sole -> 2
            | Shared -> 1
            | None -> 0))
    0 children
\end{OCaml}
  \Description{Source code} %
\caption{Deforested running example}
    \label{fig:deforestation:new_prog}
\end{minipage}
\end{figure}

\begin{figure}
  \begin{minipage}{.49\textwidth}
  \begin{align*}
    & \forall \conc 1 \leq \conc i \leq \conc {\lenvar l}, \li{\smtlib{children}'}{i} = \app f \li{\smtlib{children}}{i} \\
    & 0 + \sum_{\conc i = \conc 1}^{\conc {\lenvar l}} \li{\smtlib{children}'}{i}
  \end{align*}
  \Description{Equation} %
  \caption{\conc {\lenvar l + 1} constraints encode the original version}
    \label{fig:deforestation:og_cstr}
\end{minipage}\hfill\begin{minipage}{.49\textwidth}
  \begin{align*}
    & \phantom{forall 1 \leq i \leq \conc {\lenvar l}, \li{\smtlib{children'}}{i} = \app f \li{\smtlib{children}}{i}} \\
    & 0 + \sum_{\conc i = \conc 1}^{\conc {\lenvar{l}}} \app f \li{\smtlib{children}}{i}
  \end{align*}
  \Description{Equation} %
  \caption{One constraint encodes the deforested version}
  \label{fig:deforestation:new_cstr}
\end{minipage}
\end{figure}

\paragraph{Motivating Example}
Consider the \fold expression from the running example, recalled in
\Cref{fig:deforestation:og_prog}.
In this section, we will call $f$ the anonymous function in the \map combinator.

As explained in \Cref{sec:length},
if the length abstraction generated a length model $\conc {\lenvar l}$,
the fixed-length under-approximation will encode \ml{children} as
$\li{\smtlib{children}}1, \ldots, \li{\smtlib{children}}{\conc {\lenvar{l}}}$.
Then, it will create $\conc {\lenvar l}$ new variables
$\li{\smtlib{children}'}1, \ldots, \li{\smtlib{children}'}{\conc {\lenvar{l}}}$
to represent the mapped list,
and finally encode the expression as $\conc {\lenvar l} + 1$ constraints, as shown in \Cref{fig:deforestation:og_cstr}.
Notice however that the intermediate variables representing the map can be removed and their
definitions inlined, to yield a single constraint shown in \Cref{fig:deforestation:new_cstr}.

To generalize this kind of rewriting,
we observe that it is made possible
not only by the inlining of specific equalities,
but more fundamentally because \map and \fold compose into a single \fold expression.
In our example, we can rewrite the entire expression into the term presented in \Cref{fig:deforestation:new_prog}.

\newcommand\foldofmap{\eqref{deforestation:foldmap}\xspace}
\newcommand\mapofmap{\eqref{deforestation:mapmap}\xspace}
\newcommand\lenofmap{\eqref{deforestation:lenmap}\xspace}

\paragraph{Rewriting rules}
To reduce the size of the SMT queries and simplify reasoning,
\duelist provides the following rewriting rules:
\begin{itemize}
\item For any functions
        $f: \gamma \to \beta \to \gamma$,
        $g: \alpha \to \beta$,
        expression $i: \gamma$,
        and list expression $l: \listtype\alpha$,
        \[
            \appfold fi{(\appmap gl)}
            = \appfold
            {\big(\applambdatwo{acc}{x}{\apptwo f{acc}{(\app g x)}}\big)}
            il
            \label{deforestation:foldmap}
            \tag{\fold of \map}
          \]

          The \foldofmap rule captures the simplification described above on the running example.
          It makes the unfolded expression larger,
          but requires creating fewer SMT variables,
          and reduces the number of constraints sent to the SMT solver.

\item
        For any function
        $f: \alpha \to \beta$,
        and list expression $l: \listtype\alpha$,
        \[
            \applen {(\appmap fl)}
            = \applen l
            \label{deforestation:lenmap}
            \tag{\len of \map}
        \]

        \lenofmap leverages the fact that
        \map does not change list lengths.
        It resembles the extraction rule for \map (\Rule{Map}, \Cref{sec:length:ox}),
        but applies directly to list expressions.

\item
        For any functions
        $f: \beta \to \gamma$,
        $g: \alpha \to \beta$,
        and list expression $l: \listtype\alpha$,
        \[
            \appmap f{(\appmap gl)}
            = \appmap {(f \circ g)} l
            \label{deforestation:mapmap}
            \tag{\map of \map}
        \]

\mapofmap is especially useful in addition to the other two rules,
when applications of \map are nested inside of \len or \fold operators.

\end{itemize}

\paragraph{Application order}
Since the language of \duelist is pure,
expressions can be rewritten in any order.
In other words, the rewriting system is confluent.
For example, $\applen {\big(\appmap f{(\appmap gl)}\big)}$ can either be reduced to
$\applen l$ by applying ${\text\mapofmap}$ and ${\text\lenofmap}$, or by swapping the two rules.
In practice, we first apply the \mapofmap rule recursively to reduce the number of \map{}s.
Then, \foldofmap and \lenofmap only have to be applied once because their inner
\map expressions are already reduced.

\subsection{SMT-Solving Optimizations}
\label{sec:optimizations:technical}

As \duelist builds on top of existing, state-of-the-art SMT solvers,
we now discuss several common optimizations~\cite{DBLP:conf/osdi/CadarDE08,ayounSoteriaPLDI2026,pereiraSmtml2026}
implemented in our prototype, which aim to reduce the overall SMT solving time.

\paragraph{Incremental solving}
SMT solvers such as Z3 and CVC5 provide an incremental mode.
In this mode, the user chooses to push groups of constraints, and the state of
the solver is saved after each push.
The user can also pop groups of constraints, e.g., when a set of constraints is
unsatisfiable, or when a symbolic execution considers a different program path.
This mode makes it possible to reuse previously added constraints,
for which the SMT solver already has a model,
and only add some new constraints on top,
which amortizes solving time significantly.

In \duelist, this mode is used primarily in the feedback loop (\Cref{sec:length:feedback}) when the hints from the \uxs are added to the \oxs.
As the new constraints are added on top of the extracted length constraints to refine the current length abstraction,
the SMT solver can keep most of its previous state.

\paragraph{Pruning and simplifying spurious constraints}
The translation process induced by the length abstraction
(\Cref{sec:length}) can generate many redundant or spurious
constraints. For instance, as the translation in the \oxs erases
semantic details about \map applications (\Rule{Map}), different
applications of \map to the same list might lead to identical
constraints. Alternatively, when considering a fixed, constant
length in the \oxs, constraints related to the \len operator,
e.g., $\applen l = 10$, might become trivial.

To handle these cases, we therefore implement a preprocessing pass
which detects and removes redundant constraints, thus reducing
the size of SMT queries. This preprocessing phase also simplifies
trivial constraints, in some cases avoiding the need to query the SMT
solver altogether when the problem is trivially satisfiable or unsatisfiable.
Note that although SMT solvers already perform some simplifications,
which in theory lead to similar encodings,
it is empirically more efficient for callers to simplify problems
before querying SMT solvers \cite[Sec.~5]{pereiraSmtml2026}.

\section{Experimental Evaluation}
\label{sec:eval}

We now present our experimental evaluation of \duelist.
We first describe our implementation in \Cref{sec:eval:implem}, and describe the
set of benchmarks we assembled in \Cref{sec:eval:corpus}.
We then evaluate our solving heuristics through an ablation study in \Cref{sec:eval:ablation},
and finally compare \duelist to state-of-the-art solvers in \Cref{sec:eval:comp}.

\subsection{Implementation}
\label{sec:eval:implem}

\duelist is implemented in 4,000 lines of OCaml.
It reads input problems in the SMT-LIB extension presented in
\Cref{sec:language}.
\duelist uses Z3 4.15.2 as its backend SMT solver; communications between the
two are performed using the Z3 OCaml API.
Note however that \duelist does not rely on features specific to Z3; in future work, we plan to leverage the work of~\citet{pereiraSmtml2026}
to allow switching between multiple solvers and benefit from the built-in
optimizations of Smt.ml.

We performed all experiments on a desktop computer with an Intel Core i7-12700 processor and
128GB of DDR5 RAM, running Ubuntu 24.04.1.
To ensure accurate runtime measurements, we disabled HyperThreading, TurboBoost, and efficiency cores,
and computed averages over four runs
while distributing the load to minimize idle time for each core.
We set a query timeout of 30 seconds in our experimental setup, to account for use cases related to program analysis and verification.
For instance, symbolic execution tools~\cite{king1976symbolic} query SMT solvers thousands of times in a given program exploration.

\subsection{A Corpus of List Problems}
\label{sec:eval:corpus}

While several previous works investigated automated reasoning on list-like structures,
examples and benchmarks are scattered through different projects, and described using different
input languages. To enable a comparison with existing tools and foster future research
on list reasoning, we first assembled a corpus of \emph{parameterized} list problems by
manually curating and adapting problems from four real-world software projects and previous works.
In total, our corpus contains \evalNumFilesDuelist parametric benchmarks, separated into \evalNumFilesSat satisfiable and \evalNumFilesUnsat unsatisfiable instances, totaling \evalNumLinesDuelist lines of code using our SMT-LIB extension (comments and empty lines excluded from the line count).
Applying the parameters needed for our evaluation produces
\evalNumFilesParam \duelist files.
The benchmarks use various sorts, including integers and datatypes (tuples, records, enums).
We first explain how benchmarks are parameterized, and then describe the origin
of each benchmark.
Whenever relevant, each benchmark is encoded in satisfiable and unsatisfiable
versions;
in those cases, variants only differ on a few assertions or parameters.

\begin{figure}
  \lstdefinestyle{localstyle}{
    style=smtlib,
    basicstyle=\footnotesize\tt,
    morekeywords={define-catamorphism}
  }
  \begin{minipage}[t]{.5\textwidth}
\begin{lstlisting}[style=localstyle]
(declare-datatypes () 
  ((List
     (List_nil)
     (List_cons (destList_cons List_cons_recd)))
   (List_cons_recd
     (List_cons_recd (List_cons_recd_hd Int)
     (List_cons_recd_tl List)))))

(define-catamorphism SumListFn ((L List)) Int
  (ite (is-List_nil L) 0
    (+ (List_cons_recd_hd (destList_cons L))
       (SumListFn
       (List_cons_recd_tl (destList_cons L))))))

(declare-fun L1 () List)
(declare-fun L2 () List)

(assert (and 
  (is-List_nil L1)
  (not (= (SumListFn L2) 
    (+ (SumListFn L1) (SumListFn L2))))))

(check-sat)
\end{lstlisting}
    \Description{Source code} %
  \end{minipage}%
  \begin{minipage}[t]{.5\textwidth}
\begin{lstlisting}[style=localstyle]
(define-fun SumListFn ((L (List Int))) Int
  (List.fold
    (lambda ((acc Int) (x Int)) Int (+ acc x))
    0 L))

(define-fun isnil ((L (List Int))) Bool
  (= (List.len Int L) 0))

(declare-const L1 (List Int))
(declare-const L2 (List Int))

; Parametric assertions to bound list lengths.
; N will be replaced by an integer.
;param:(assert (<= (List.len Int L1) N))
;param:(assert (<= (List.len Int L2) N))

(assert (and
  (isnil L1)
  (not (= (SumListFn L2)
    (+ (SumListFn L1) (SumListFn L2))))))

(check-sat)
\end{lstlisting}
    \Description{Source code} %
    \vfill
  \end{minipage}
  \caption{Comparison of RADA source code (left)
     and the corresponding \duelist version (right)}
  \label{fig:rada}
\end{figure}

\paragraph{Benchmark parameterization}
All problems are parametric in the length of the lists they involve,
to enable an analysis of solving time with respect to the problem size.
We denote the length of the list as $N$ in the remainder of this section.
For satisfiable problems, the parameter is a lower bound on list lengths,
to force larger problems even when smaller lists would satisfy the constraints.
For unsatisfiable problems, the parameter is an upper bound on list lengths,
to bound the search space.%

\paragraph{RADA~\cite{DBLP:journals/jar/PhamGW16}}
RADA is a solver specializing on algebraic datatypes.
It contains two variations of benchmarks with list problems.
Those lists are encoded as algebraic data structures, with a head and a tail.
Then, RADA problems describe constraints using catamorphisms~\cite{DBLP:conf/fpca/MeijerFP91}, a generalized \fold operation on algebraic datatypes.
We encode a list sum problem (\bench{Checksum}), as well as a problem where list elements
must alternate between cases of an enum (\bench{Odd-Even}).

To illustrate our benchmark writing process,
we compare in \Cref{fig:rada}
a shortened version of the original RADA \bench{Checksum} problem
with the \duelist benchmark inspired by it.
The RADA code first defines the recursive list datatype
and a catamorphism that recursively sums the elements of a list,
before encoding an unsatisfiable assertion about two lists
\smtlib{L1} and \smtlib{L2}: if \smtlib{L1} is empty,
adding its sum to the sum of \smtlib{L2} should be neutral.
The \duelist code, on the other hand,
defines the sum using its native \smtlib{List} type and \fold combinator.
However, checking if \smtlib{L1} is empty
requires an additional function \smtlib{isnil},
that uses the \len combinator.
Finally, to guarantee termination and make the benchmark parametric,
the \duelist code bounds the lengths of \smtlib{L1} and \smtlib{L2}
with an integer parameter \smtlib{N}.

\paragraph{Solver Competitions}
The SyGuS competition~
\cite{DBLP:journals/corr/abs-1711-11438,DBLP:journals/cacm/AlurSFS18}
focuses on syntax-guided synthesis problems.
While these problems aim at synthesizing functions that satisfy their
constraints, some can be seen as arrays (arrays and functions are equivalent in
SMT-LIB), or can be formulated easily with lists.
Thus, we create problems inspired by SyGuS Comp problems,
where list combinators could be used.
We encode array problems with \ml{exists}, \ml{foldi}, \ml{nth}, or \ml{map} operations,
formulated in terms of invariants and transformations,
by using our \smtlib{fold} and \smtlib{map} (\bench{Simple array}, \bench{indexOf}, \bench{Contains}, \bench{ExistsSame}, \bench{Swap01}, \bench{FindMax}).
We also encode an ``agent'' problem, called the \bench{Taxi dropoff} problem.
A taxi has to go from a source to a target point on a grid while avoiding obstacles.
The SyGuS problem asks for a function that gives the taxi a direction at any
point on the grid, whereas we try to find a list of directions that it should follow in order.

We surveyed benchmarks from the SMT-LIB library~
\cite{zenodo:SMT-LIB/2024/non-incremental},
in the single-query QF\_DT, UFDT, UFDTLIA, and AUFDTLIA tracks.
These benchmarks are used to compare SMT solvers
in the SMT-COMP competition~\cite{SMTCOMP2024},
and are usually generated automatically
from verification goals for other tools,
or randomly from parametric problems.
However, QF\_DT does not allow recursion,
and the problems that do allow quantifiers and encode list-like structures
aim mainly at proving inductive properties of functions
like \smtlib{append} or \smtlib{rev}.
\duelist, on the other hand,
targets problems from program verification tools,
such as path conditions generated by symbolic execution,
that require synthesizing lists of arbitrary sizes
constrained by \map or \fold combinators.

\paragraph{OCaml~\cite{leroy2024ocaml}}
We encode several problems inspired by the OCaml compiler test suite.
While the original test suite consists of assertions comparing the result of the
functions to their expected values, we instead query \duelist for a model of lists that satisfy constraints.
We encode functions from the OCaml standard library that can be written as
\fold{}s: \ml{exists} (\bench{Exists}), \ml{nth} (\bench{Nth}), \bench{\ml{List.hd}}, \bench{\ml{List.filter}}, \bench{\ml{List.find_map}}, and checking that a list is sorted (\bench{Sorted}).

Note that the \bench{\ml{List.filter}} benchmark
does not require a proper \filter combinator.
Instead, we show that when a \filter{}ed list is used as input in a $\app\fold f$,
$\appfilter{p}{l}$ can be encoded using an Option type, by
first mapping $(\applambda x{\appite{(\app px)}
    {(\app{\smtlib{Some}}x)}
    {\smtlib{None}}})$
on $l$, and then applying
$$\app\fold{(\applambdatwo{acc}{x}{\appmatch{x}
\;| \smtlib{None} \to acc
\;| \app{\smtlib{Some}}{x'} \to \apptwo{f}{acc}{x'}})}.$$

\paragraph{Catala~\cite{DBLP:journals/pacmpl/MerigouxCP21}}
We extract several list-specific benchmarks from existing programs written in the domain-specific language Catala.
Catala programs implement a range of legislative texts focusing on computational law (e.g., taxes, or social benefits),
and heavily rely on lists, e.g., to represent the children in a household, or to capture the properties owned by an individual.
The Catala language possesses specific program constructs which aim to encode the structure and logic of the law;
we manually identify list-intensive computations, and translate the code to make it compatible with \duelist's language.
\bench{Running example} and \bench{Brackets} reflect computations encoded in the French social
benefits.
The first program computes a children custody score, used to determine eligibility to family benefits,
while the second example computes income brackets for housing benefits.
\bench{Inheritance law} specifies whether a list of children can inherit
according to French inheritance laws.

\paragraph{Custom problems}
To evaluate the impact of our optimizations, we also write custom problems
stress-testing different patterns targeted by our heuristics
(\bench{\len of \map}, \bench{\fold of \map of \map}, \bench{Neutral}),
and testing specific list shapes (\bench{All 2}, \bench{Two lists}).

\subsection{Ablation Study}
\label{sec:eval:ablation}

\begin{table}
    \begin{tabular}{lrrrrr}
        \toprule
        Benchmark                 &    N &     No optim. &   Deforestation &   Summarization &    All optim. \\
        \midrule
        \len of \map            & 5000 &  0.42s ± 0.00 &    0.05s ± 0.00 &    0.42s ± 0.00 &  0.10s ± 0.00 \\
        Neutral                   & 3000 &  4.79s ± 0.04 &    4.55s ± 0.05 &    0.05s ± 0.00 &  0.11s ± 0.00 \\
        \fold of \map of \map  & 3000 &  9.08s ± 0.06 &    4.73s ± 0.04 &    8.95s ± 0.05 &  0.11s ± 0.00 \\
        Contains                  & 3000 &  0.92s ± 0.00 &    0.92s ± 0.01 &    0.05s ± 0.00 &  0.11s ± 0.00 \\
        Nth                       &  100 &  0.02s ± 0.00 &    0.02s ± 0.00 &    0.02s ± 0.00 &  0.03s ± 0.00 \\
        Two lists                 &  100 &  0.02s ± 0.00 &    0.02s ± 0.00 &    0.02s ± 0.00 &  0.03s ± 0.00 \\
        ExistsSame                & 5000 &  4.38s ± 0.06 &    3.90s ± 0.10 &    4.29s ± 0.02 &  3.85s ± 0.02 \\
        Checksum                  &  100 &  0.02s ± 0.00 &    0.02s ± 0.00 &    0.02s ± 0.00 &  0.03s ± 0.00 \\
        Simple array              &  100 &  0.03s ± 0.00 &    0.03s ± 0.00 &    0.03s ± 0.00 &  0.03s ± 0.00 \\
        Inheritance law           &  100 &  0.03s ± 0.00 &    0.03s ± 0.00 &    0.03s ± 0.00 &  0.04s ± 0.00 \\
        List.find\_map             & 5000 &  3.00s ± 0.04 &    2.83s ± 0.03 &    2.94s ± 0.03 &  3.14s ± 0.04 \\
        Swap01                    & 5000 &  4.10s ± 0.06 &    3.56s ± 0.04 &    4.18s ± 0.02 &  4.28s ± 0.04 \\
        Sorted                    & 3000 &  5.24s ± 0.07 &    5.26s ± 0.03 &    5.21s ± 0.04 &  5.37s ± 0.13 \\
        List.hd                   & 5000 &  0.65s ± 0.01 &    0.65s ± 0.00 &    0.66s ± 0.00 &  0.75s ± 0.00 \\
        Odd-Even                  & 2000 &  4.63s ± 0.03 &    4.38s ± 0.08 &    4.45s ± 0.09 &  4.28s ± 0.03 \\
        Exists                    &   70 &  0.06s ± 0.00 &    0.06s ± 0.00 &    0.06s ± 0.00 &  0.08s ± 0.00 \\
        Running example           &   20 &  0.55s ± 0.00 &    0.50s ± 0.00 &    0.51s ± 0.00 &  0.53s ± 0.00 \\
        Brackets                  &   50 &         $\infty$ &           $\infty$ &           $\infty$ &  0.54s ± 0.01 \\
        All 2                     &  100 &  0.62s ± 0.01 &    0.60s ± 0.00 &    0.60s ± 0.00 &  0.61s ± 0.00 \\
        FindMax                   & 1000 &  0.67s ± 0.00 &    0.68s ± 0.00 &    0.69s ± 0.00 &  0.71s ± 0.00 \\
        Taxi dropoff              &   20 &  0.57s ± 0.00 &    0.46s ± 0.00 &    0.46s ± 0.00 &  0.77s ± 0.00 \\
        indexOf                   &  100 & 17.63s ± 0.02 &   17.72s ± 0.02 &   17.64s ± 0.02 & 17.68s ± 0.09 \\
        List.filter               &   50 &  0.30s ± 0.00 &    0.04s ± 0.00 &    0.30s ± 0.00 &         $\infty$ \\
        \midrule
        \len of \map & 5000 &  0.01s ± 0.00 &    0.01s ± 0.00 &    0.01s ± 0.00 &  0.01s ± 0.00 \\
        List.hd                   &  500 & 18.69s ± 0.08 &   18.92s ± 0.03 &   18.72s ± 0.05 &  0.01s ± 0.00 \\
        Checksum                  &  500 &  9.79s ± 0.01 &    9.75s ± 0.05 &    0.01s ± 0.00 &  0.02s ± 0.00 \\
        All 2                     &   30 &  0.12s ± 0.00 &    0.12s ± 0.00 &    0.12s ± 0.00 &  0.12s ± 0.00 \\
        Swap01                    &  300 & 10.76s ± 0.08 &    5.22s ± 0.07 &   10.81s ± 0.08 &  5.22s ± 0.01 \\
        Nth                       &   40 &  0.13s ± 0.00 &    0.12s ± 0.00 &    0.12s ± 0.00 &  0.13s ± 0.00 \\
        Simple array              &   30 &  0.16s ± 0.00 &    0.15s ± 0.00 &    0.16s ± 0.00 &  0.15s ± 0.00 \\
        Exists                    &   40 &  0.64s ± 0.01 &    0.64s ± 0.00 &    0.65s ± 0.01 &  0.66s ± 0.00 \\
        \fold of \map of \map  &  200 &  5.44s ± 0.01 &    1.22s ± 0.01 &    5.31s ± 0.06 &  1.24s ± 0.02 \\
        Running example           &  100 &  2.57s ± 0.00 &    2.08s ± 0.01 &    2.56s ± 0.01 &  2.05s ± 0.00 \\
        FindMax                   &   30 &  3.44s ± 0.01 &    3.65s ± 0.00 &    4.04s ± 0.00 &  3.33s ± 0.00 \\
        List.find\_map             &  100 &  3.93s ± 0.02 &    3.78s ± 0.00 &    3.90s ± 0.00 &  3.81s ± 0.00 \\
        Sorted                    &  200 &  5.52s ± 0.02 &    5.56s ± 0.03 &    5.68s ± 0.11 &  5.59s ± 0.01 \\
        Taxi dropoff              &   20 & 12.69s ± 0.05 &   13.14s ± 0.04 &   12.65s ± 0.02 & 12.54s ± 0.01 \\
        indexOf                   &  100 & 17.94s ± 0.03 &   18.04s ± 0.04 &   17.92s ± 0.02 & 17.83s ± 0.04 \\
        List.filter               &   30 &  3.23s ± 0.01 &    2.58s ± 0.00 &    3.10s ± 0.01 &         $\infty$ \\
        \bottomrule
    \end{tabular}

    \caption{Comparison between \duelist's optimizations.
        Average time on four runs with problem size N,
        and standard deviation displayed after ±.
        $\infty$ denotes a timeout (30s).
        Satisfiable problems are shown in the top part of the table,
        unsatisfiable problems follow in the bottom part.
    }
    \label{table:ablation}
\end{table}

We compare in \Cref{table:ablation} the total solving time for each problem with no optimizations,
with individual optimizations, and with all optimizations active.
The two individual optimizations are
the deforestation (\Cref{sec:optimizations:rewrite}),
and list summarization (\Cref{sec:fold:neutral}),
and the reachability heuristic (\Cref{sec:optimizations:reachability})
is active when all optimizations are.
Well-established SMT optimizations (\Cref{sec:optimizations:technical}) are always enabled, and included in our baseline.

We choose the specific parameter $N$ for each problem
so that it is large enough to measure statistically significant runtimes
when all optimizations apply.

On problems that compose combinators, like \bench{\len of \map} or \bench{List.filter},
the deforestation optimization speeds up solving time significantly.
This effect is especially visible on unsatisfiable problems like \bench{Swap01},
because the List encoder generates a large number of constraints.
The summarization optimization does not apply very often on its own,
except for problems with few list constraints, like SyGuS' \bench{Contains}.
Nonetheless, we observe on examples like our \bench{\fold of \map of \map} and SyGuS' \bench{Swap01} that deforesting the constraints first can help trigger summarization.
Moreover, we observe that summarization trivializes the unsatisfiable problem \bench{Checksum} from RADA.
In this example, the summarization interacts with the length abstraction to transform a list assertion in an arithmetic problem,
that Z3 proves unsatisfiable quickly.
Similarly, \bench{List.hd} is trivialized by the reachability heuristic
when all optimizations are active.

The optimizations have no effect on a few of the longer-running benchmarks,
because their shape does not allow optimizations to apply.
For instance, \bench{indexOf} is hard because it requires that
a specific value be present at a specific index in a list,
but it is not written with a deep composition of combinators.
\bench{Taxi dropoff} is essentially a path-finding problem,
and makes heavy use of algebraic datatypes,
but its solutions cannot be summarized efficiently by our current optimizations.
Nonetheless, \duelist's modular architecture
allows adding new heuristics in the under- and over-approximating solvers,
like those of \Cref{sec:fold},
to target specific issues in the future.

Finally, we observe that the combination of optimizations makes \bench{List.filter} time out,
although deforestation does reduce its solving time.
In that case, deforestation allows summarization to trigger,
but Z3 times out while solving the summarization's condition (\Cref{eqn:neutral}).
In practice, \duelist would be configured to give up the optimization after a short solving time.

\subsection{Comparison with State-of-the-art Solvers}
\label{sec:eval:comp}

Since \duelist tackles input problems written in a custom extension of
SMT-LIB, it cannot be directly compared with other state-of-the-art solvers.
To evaluate our approach against existing tools, we implement an automated translation of \duelist's
language into input languages for other solvers,
and three different theories: datatypes, arrays, and sequences.
Note that although \duelist's theory of lists,
and its internal queries to Z3, are quantifier-free,
these encodings require universal quantifiers, or alternatively,
recursive functions, to encode recursive definitions in other formats.

Our benchmark corpus
is expanded into \evalNumFilesParam \duelist benchmarks by varying the parameters,
then translated into the 3 supported theories,
producing \evalNumFilesGenerated files
totaling \evalNumLinesGenerated lines of SMT-LIB.

\paragraph{Theory of datatypes}
Using the standard theory of datatypes,
we declare a list datatype consisting of
the usual \smtlib{nil} and \smtlib{cons} constructors.
\begin{SMT-LIB}
    (declare-datatype DTList (par (T) ((nil) (cons (head T) (tail (DTList T))))))
\end{SMT-LIB}
The \fold, \map and \len combinators are defined as recursive functions.
Note that although the \smtlib{DTList} type is parametric
in the type \smtlib{T} of its elements,
these functions must be defined explicitly for each \smtlib{T}
in the benchmark.
For instance, for \smtlib{(DTList Int)},
and functions \smtlib{f} and \smtlib{g}:
\begin{SMT-LIB}
    (define-fun-rec len ((l (DTList Int))) Int (ite (= l nil)) 0 (+ 1 (len (tail l))))
    (define-fun-rec map_f ((l (DTList Int))) (DTList Int)
        (ite (= l nil) nil (cons (f (head l)) (map_f (tail l)))))
    (define-fun-rec fold_g ((acc Int) (l (DTList Int))) Int
        (ite (= l nil) acc (fold_g (g acc (head l)) (tail l))))
\end{SMT-LIB}

\paragraph{Theory of arrays}
Using the standard theory of arrays,
we define a list as a record, consisting of
an array indexed over integers and
an integer representing the length of the list.
\begin{SMT-LIB}
    (declare-datatype AList (par (T) ((alist (length Int) (array (Array Int T))))))
\end{SMT-LIB}
Using this encoding, \len simply looks up the length of the record:
\smtlib{(length l)}.
Since SMT-LIB arrays can also be viewed as functions,
we define $\app\map f$ as the composition of $f$ with the array,
using a \smtlib{lambda}-expression.
\begin{SMT-LIB}
    (lambda ((i Int)) (f (select (array l) i)))
\end{SMT-LIB}
Finally, $\app\fold f$ is defined as a recursive function that
traverses the array by increasing indices, until the length is reached.
\begin{SMT-LIB}
    (define-fun-rec fold_f ((k Int) (acc T1) (l (AList T2))) T1
        (ite (= k (length l)) acc (fold_f (+ k 1) (f acc (select (array l) k)) l)))
    ; call: (fold_f 0 init)
\end{SMT-LIB}

\paragraph{Theory of sequences}
Our encoding of lists as sequences relies on the
Z3-specific interface added in version 4.9.0 (2022)
for the theory of sequences, which is not standardized.
We use it for its unique combinators
\smtlib{seq.fold}, \smtlib{seq.map} and \smtlib{seq.len},
that use the \smtlib{lambda} keyword the same way \duelist does.
\begin{SMT-LIB}
    (seq.len l)
    (seq.map (lambda ((x Int)) (f x)) l)
    (seq.fold_left (lambda ((acc Int) (x Int)) (g acc x)) init l)
\end{SMT-LIB}

\paragraph{Comparison with RADA}
Even though some of our benchmarks were adapted from RADA
\cite{DBLP:journals/jar/PhamGW16}, we have been unable to reproduce RADA's results locally
and therefore do not include this tool in our comparison.
Indeed, even when using the original RADA format, we observed that RADA
answered \SAT on several of its problems that are, in fact, unsatisfiable.
We also ran RADA on our benchmarking suite to get partial results,
but we found that it answered \SAT on every problem, even unsatisfiable ones.
Moreover, it did so in constant time ($\simeq 20$ms), regardless of the problem.
This suggests that, despite our best effort, and after using
the recommended Z3 and CVC4 versions from RADA's release year (2013),
we could not reproduce RADA's original running environment,
which was not preserved in an artifact.

\paragraph{Comparison with CVC5}
We ran CVC5 version 1.3.4 on our datatype encoding,
using its finite model finding option~
\cite{DBLP:conf/cade/ReynoldsBCT16}
to handle recursive definitions.
Although CVC5 provides a theory of sequences,
it does not currently implement the
\smtlib{seq.fold} and \smtlib{seq.map} combinators available in Z3,
so we cannot compare its performance on this encoding.
Likewise, while the higher-order logic of CVC5 (HO\_ALL) supports
the \smtlib{lambda} keyword introduced in version 2.7 of the SMT-LIB standard,
it disallows defining arrays as anonymous functions with this keyword like Z3 does,
so we cannot compare CVC5 on the array encoding.

Also note that, while we experimented with several solvers that participate
in datatypes tracks of SMT-COMP
(Algaroba~\cite{DBLP:conf/aaai/ShahMS24}, CVC5, SMTInterpol~\cite{DBLP:conf/spin/ChristHN12}),
we only included the solver that answered correctly in our time budget
on at least one benchmark.
Indeed, Algaroba is a quantifier-free solver,
and although SMTInterpol allows quantifiers,
it is not specialized in datatypes, and it timed out on all of our benchmarks.

\paragraph{Comparison with SyGuS solvers}
Although some of our benchmarks were adapted from SyGuS Comp problems,
we cannot meaningfully compare \duelist with SyGuS-capable solvers
on our benchmarks,
because function synthesis is a fundamentally different class of problems.
Indeed, our benchmarks would not take advantage from SyGuS grammars,
and instead require synthesizing complex list constants,
for which synthesizers need specialized algorithms~
\cite{abateSynthesisingProgramsNontrivial2023}.
To confirm this intuition,
we tried running CVC5 in SyGuS mode on our datatype encoding,
and observed no difference with our CVC5 runs.

\begin{table}
    \begin{tabular}{lrrrrr}
        \toprule
        Benchmark                 &   N &          \duelist &           Z3 (DT) &          Z3 (Seq) &        Z3 (Array) \\
        \midrule
        \len of \map            & 100 &  \bf 0.01s ± 0.00 &      1.33s ± 0.01 &      0.40s ± 0.00 & \bf 0.01s ± 0.00 \\
        Neutral                   & 100 &  \bf 0.02s ± 0.00 &      0.99s ± 0.00 &      0.49s ± 0.01 &      0.19s ± 0.00 \\
        \fold of \map of \map  & 100 &  \bf 0.02s ± 0.00 &      6.11s ± 0.03 &      0.52s ± 0.00 &         (unknown) \\
        Contains                  & 200 &  \bf 0.02s ± 0.00 &      2.75s ± 0.01 &      4.85s ± 0.05 &      1.37s ± 0.01 \\
        Nth                       & 100 &  \bf 0.03s ± 0.00 &      1.15s ± 0.01 &      5.08s ± 0.47 &             $\infty$ \\
        Two lists                 & 100 &  \bf 0.03s ± 0.00 &      3.89s ± 0.02 &      1.33s ± 0.01 &      0.72s ± 0.01 \\
        ExistsSame                & 100 &  \bf 0.03s ± 0.00 &      2.47s ± 0.04 &      2.22s ± 0.06 &         (unknown) \\
        Checksum                  & 100 &  \bf 0.03s ± 0.00 &      3.47s ± 0.01 &      1.02s ± 0.01 &      0.20s ± 0.00 \\
        Simple array              & 100 &  \bf 0.03s ± 0.00 &      4.84s ± 0.01 &      0.54s ± 0.01 &         (unknown) \\
        Inheritance law           & 100 &  \bf 0.04s ± 0.00 &      2.68s ± 0.01 &      0.59s ± 0.00 &         (unknown) \\
        List.find\_map             & 100 &  \bf 0.05s ± 0.00 &      1.70s ± 0.01 &      1.48s ± 0.04 &         (unknown) \\
        Swap01                    & 100 &  \bf 0.05s ± 0.00 &      4.21s ± 0.07 &      0.52s ± 0.01 &         (unknown) \\
        Sorted                    & 200 &  \bf 0.05s ± 0.00 &     26.70s ± 0.15 &      9.25s ± 0.32 &      9.53s ± 0.13 \\
        List.hd                   & 300 &  \bf 0.06s ± 0.00 &             $\infty$ &     18.41s ± 0.63 &      1.15s ± 0.01 \\
        Odd-Even                  & 100 &  \bf 0.06s ± 0.00 &             $\infty$ &      4.92s ± 0.01 &             $\infty$ \\
        Exists                    &  70 &  \bf 0.08s ± 0.00 &     25.89s ± 0.09 &      0.27s ± 0.00 &             $\infty$ \\
        Running example           &  20 &       0.53s ± 0.00 & \bf 0.26s ± 0.00 &             $\infty$ &         (unknown) \\
        Brackets                  &  50 &  \bf 0.54s ± 0.01 &             $\infty$ &             $\infty$ &             $\infty$ \\
        All 2                     & 100 &  \bf 0.61s ± 0.00 &      5.31s ± 0.00 &             $\infty$ &      0.68s ± 0.01 \\
        FindMax                   & 200 &  \bf 0.10s ± 0.00 &             $\infty$ &      4.70s ± 0.23 &             $\infty$ \\
        Taxi dropoff              &  20 &       0.77s ± 0.00 & \bf 0.49s ± 0.00 &      6.35s ± 0.02 &      2.19s ± 0.02 \\
        indexOf                   & 100 &      17.68s ± 0.09 & \bf 0.91s ± 0.00 &             $\infty$ &             $\infty$ \\
        List.filter               &  50 &              $\infty$ & \bf 0.64s ± 0.01 &      2.48s ± 0.05 &         (unknown) \\
        \midrule
        \len of \map & 100 &  \bf 0.01s ± 0.00 &             $\infty$ & \bf 0.01s ± 0.00 & \bf 0.01s ± 0.00 \\
        List.hd                   &  50 &  \bf 0.01s ± 0.00 &             $\infty$ &      3.15s ± 0.04 &             $\infty$ \\
        Checksum                  & 500 &       0.02s ± 0.00 &             $\infty$ & \bf 0.01s ± 0.00 & \bf 0.01s ± 0.00 \\
        All 2                     &  30 &  \bf 0.12s ± 0.00 &             $\infty$ &      0.72s ± 0.01 &             $\infty$ \\
        Swap01                    &  30 &  \bf 0.13s ± 0.00 &             $\infty$ &      1.35s ± 0.01 &         (unknown) \\
        Nth                       &  40 &  \bf 0.13s ± 0.00 &             $\infty$ &      3.83s ± 0.01 &             $\infty$ \\
        Simple array              &  30 &  \bf 0.15s ± 0.00 &             $\infty$ &      1.05s ± 0.01 &             $\infty$ \\
        Exists                    &  40 &  \bf 0.66s ± 0.00 &             $\infty$ &      8.47s ± 0.02 &             $\infty$ \\
        \fold of \map of \map  &  50 &  \bf 0.12s ± 0.00 &             $\infty$ &      1.80s ± 0.03 &             $\infty$ \\
        Running example           & 100 &  \bf 2.05s ± 0.00 &             $\infty$ &             $\infty$ &             $\infty$ \\
        FindMax                   &  30 &       3.33s ± 0.00 &             $\infty$ &     14.18s ± 0.05 & \bf 0.01s ± 0.00 \\
        List.find\_map             & 100 &  \bf 3.81s ± 0.00 &             $\infty$ &             $\infty$ &             $\infty$ \\
        Sorted                    &  50 &  \bf 0.41s ± 0.00 &             $\infty$ &     27.70s ± 0.54 &             $\infty$ \\
        Taxi dropoff              &  20 & \bf 12.54s ± 0.01 &             $\infty$ &             $\infty$ &             $\infty$ \\
        indexOf                   & 100 &      17.83s ± 0.04 &             $\infty$ &             $\infty$ & \bf 0.03s ± 0.00 \\
        List.filter               &  30 &              $\infty$ &             $\infty$ & \bf 2.12s ± 0.01 &             $\infty$ \\
        \bottomrule
    \end{tabular}

    \caption{Comparison between \duelist and state of the art solvers.
        Average time on four runs with problem size N,
        and standard deviation displayed after ±.
        $\infty$ denotes a timeout (30s).
        The best time for each benchmark is in bold.
        Satisfiable problems are shown in the top part of the table,
        unsatisfiable problems follow in the bottom part.
    }
    \label{table:compare}
\end{table}

\paragraph{Comparison table}
We therefore compare \duelist to Z3 with the datatype, sequence and array encodings,
and to CVC5 with the datatype encoding.
We provide the results of our evaluation in \Cref{table:compare},
except for those of CVC5 with the datatype encoding,
since it times out on all but one benchmark.
We choose the parameter $N$ for each problem to be as large as possible
while avoiding timeouts in the instances of Z3.
Empirically, \duelist outperforms existing tools in the vast majority of the cases,
both on satisfiable and unsatisfiable instances.
We also observe that the datatype encoding solves most satisfiable problems,
although at a slower pace than \duelist.
It struggles, however, with unsatisfiable problems: we suspect this might be because it is
unable to leverage the upper bound on the state space and never terminates.
On satisfiable problems, the sequence encoding performs similarly to the
datatype encoding, and is slower on problems that rely heavily on datatypes (\bench{Taxi Dropoff}, \bench{Running example}).
On unsatisfiable problems, it can establish unsatisfiability in most cases,
but \duelist answers up to 315 times faster (\bench{List.hd}).
The array encoding is the most efficient of all encodings, but it seldom works.
In particular, it outperforms \duelist
on unsatisfiable problems that rely heavily on indices
(\bench{indexOf}).
Z3 is able to handle the unsatisfiable \bench{Checksum} benchmark
with the sequence and array encodings:
using their efficient encoding for \len,
it performs a symbolic unification that trivializes the unsatisfiability proof.
Also note that \bench{Taxi dropoff},
where no \duelist optimizations
and no specialized theories (such as arrays) apply,
is still solved by \duelist in the unsatisfiable case,
while state-of-the-art solvers cannot.

Finally, CVC5~(DT) finishes in 10.11s ± 0.19
on the satisfiable \bench{Taxi dropoff} problem with $N=20$,
the same order of magnitude as Z3~(Seq).
While this is the only problem where CVC5 does not time out
with the parameters of \Cref{table:compare},
its performance on smaller problems
is presented in the scalability evaluation below.

\begin{figure}
    \includegraphics[width=\textwidth]{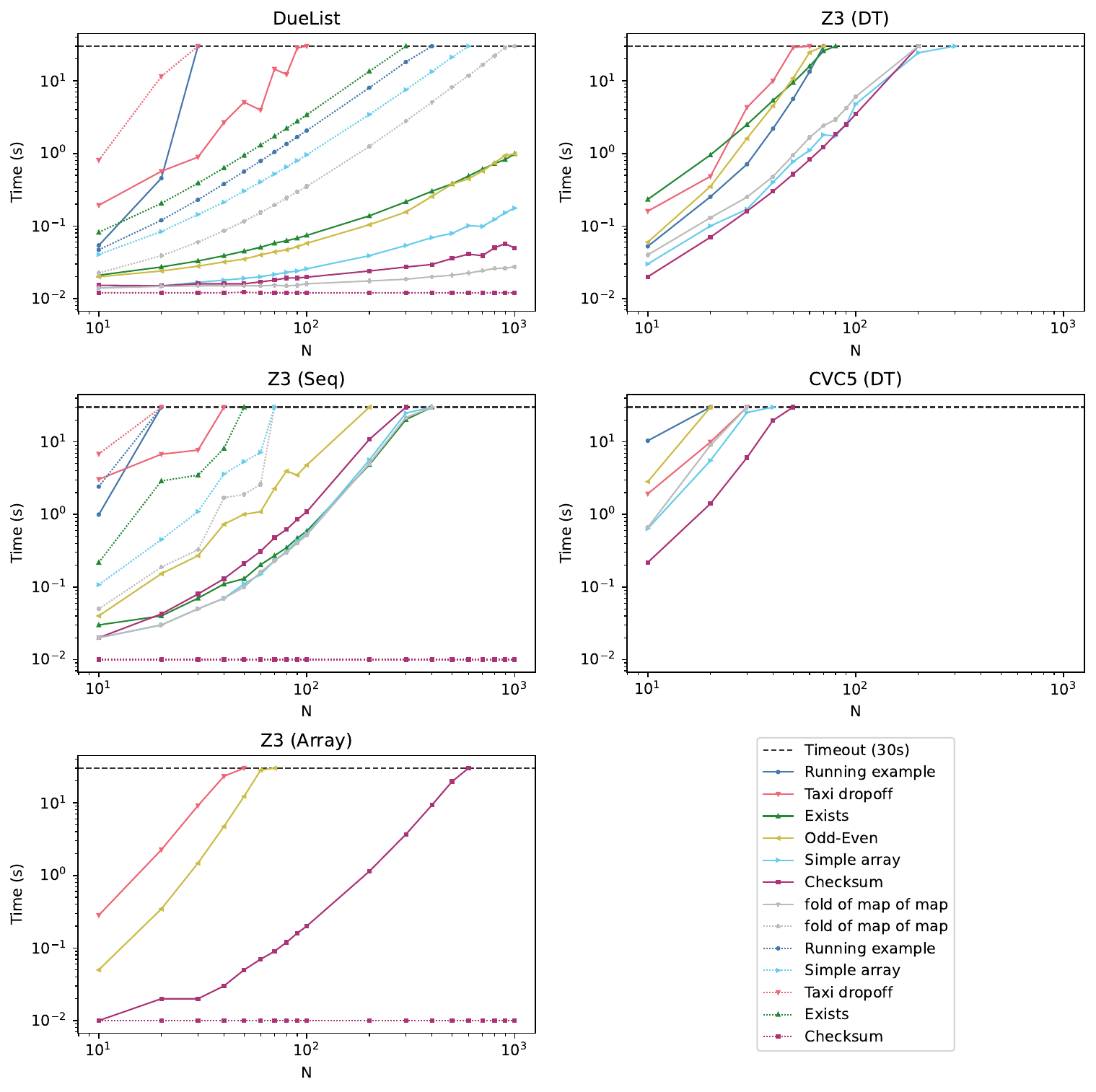}
    \Description{Figure with 5 subplots, one for each solver. The results are described in the corresponding paragraph.}
    \caption{Solving time (average of four runs) vs problem size,
        ranging from N=10 to N=1000.
        Times after the first timeout (30s) are not drawn.
      }
    \label{fig:graphs}
\end{figure}

\paragraph{Evaluating scalability}
To evaluate the scalability of \duelist compared to other solvers, we
now leverage the parametric nature of our benchmarks.
We run each problem with parameters ranging from N = 10 to N = 1000 and plot
the results in \Cref{fig:graphs}.
To avoid cluttering the plots, we only show a subset of the problems.
Preference was given to problems that Z3 and CVC5 could also solve, problems from the
literature, and problems that show the impact of \duelist's heuristics.
Unsatisfiable problems are drawn in dotted lines,
and all plots share the same legend and log-log scale.
For instance, \duelist solves the unsatisfiable \bench{Simple Array} benchmark
(cyan arrow to the right, dotted)
with N=30 in 0.1s, while Z3 (Seq) solves it in 1s, and Z3 (Array) times out.

As expected, we remark that unsatisfiable problems are more expensive than
most satisfiable ones.
As mentioned previously, \bench{Checksum} is trivialized by a Z3 unification pass
with the sequence and array encoding.
On unsatisfiable problems, only the Z3 sequence encoding can provide actionable
results, but only up to a problem size of N = 70.
Within the same time budget, \duelist can solve instances at least four times
larger.

On satisfiable problems, the array encoding scales reasonably well on
specialized instances.
The sequence and datatype encodings are able to solve more satisfiable
problems, but they require at least one order of magnitude more time than \duelist in all
cases except the \bench{Running example}.
On the unsatisfiable version of the \bench{Running example},
\duelist is the only solver to answer if $N \ge 20$.
However, the datatype encoding allows Z3 to perform slightly better
on the satisfiable version.
We believe that this is partly due to the fact that
this benchmark requires many extra constraints to generate useful lists,
which in turn creates more intricate SMT constraints
that rely heavily on datatypes.
Using the same encoding, CVC5 solves smaller benchmarks than Z3,
and requires up to two orders of magnitude more time.

\section{Related work}

\looseness=-1
We start by surveying relevant related SMT theories.
Historically, the array theory has been standardized first in SMT-LIB, followed by string theory.
Algebraic datatypes have then been added in the SMT-LIB language, and theories around sequences are being integrated in various solvers.

\paragraph{Arrays}
Following the SMT-LIB standard, arrays are modeled with a \smtlib{store} and a \smtlib{select} function
\cite{demouraGeneralizedEfficientArray2009}\cite[Fig.~3.3]{SMT-LIBv2.7}.
They can be indexed by any type (in particular, nested arrays are supported), and have elements of any type, making them
essentially indistinguishable from functions.
Because of this, arrays do not have an intrinsic length, and constraints must
explicitly state to which subset of the domain they apply.
Our list language exposes a higher-level interface, which may help in proving
more of the considered benchmarks \UNSAT.
Nevertheless, our list language can encode \smtlib{store} and \smtlib{select}
functions through a \fold operating on a tuple.
While \duelist essentially acts as a functor transforming a solver operating
over a theory $\mathcal{T}$ into a solver operating on
$\textit{List}[\mathcal{T}]$, we have not explored the capabilities of our approach
for nested lists.
\citet{DBLP:conf/sac/LeinoM09} describe an axiomatization of common array comprehensions used within a deductive verifier to discharge automatically proofs to SMT solvers.
\citet{DBLP:conf/cade/HerrmannR25} study complexity classes of the array theory on constant arrays and sum constraints.
These considered cases can be encoded within \duelist, as comprehensions can be rewritten as \fold operations.

\paragraph{Strings}
SMT-LIB introduced in 2018 a string theory~\cite{smtlib/changelog},
where strings are first-class elements, and not built recursively.
This theory cannot be leveraged directly to get a theory of lists.
First, strings are made of letters that come from a finite alphabet,
which is enough to represent ASCII or Unicode characters,
but not the infinite set of integers.
Additionally, operations on strings,
such as concatenation, slicing, or regular expression matching,
usually rely on the structure of the string.
The only operation on a string's \emph{contents} in SMT-LIB's theory
\cite{SMT-LIB/theories/Strings} consists in replacing a substring.
In particular, string theories do not define operations resembling a \fold.
However, string solvers like Z3str4~\cite{moraZ3str4MultiarmedString2021} can
efficiently handle strings of symbolic lengths with the help of a \emph{Length
  Abstraction Solver}: this is similar to \duelist's design to support lists
of symbolic lengths.

\paragraph{Algebraic datatypes}%
SMT-LIB version 2.6 introduced native support for user-defined
algebraic datatypes, allowing an inductive encoding of lists through a \ml{nil} constructor and
a \ml{cons} or \ml{concat} constructor.
Most operations over lists can then be defined recursively, as in functional languages.
This encoding enables writing constraints on lists of symbolic size, as well as constraints on the contents of these lists.
\citet{DBLP:conf/cade/ReynoldsB15} describe a decision procedure for
(co)datatypes implemented within CVC4.
\citet{DBLP:journals/jar/PhamGW16} introduce an unrolling-based decision
procedure for \emph{catamorphisms}, which can be seen as a generalization of
\fold for algebraic datatypes.
\citet{kobayashiSolvableTuplePatterns2026}
introduce solvable tuple patterns
to solve recursive list-like data structures in Constrained Horn Clauses solvers,
but this formalism cannot currently express \map or \fold combinators.
Since the recursive datatype encoding requires quantifiers,
it exceeds the theories supported by
quantifier-free datatype solvers like Algaroba~\cite{DBLP:conf/aaai/ShahMS24}.
In our experiments, we found that SMT solvers are usually unable to prove \UNSAT
even on problems with a bounded state-space to explore, suggesting they do not
incorporate a generalization of a length abstraction solver.

\paragraph{Other SMT-based approaches.}
Solvers such as CVC5 or Z3 define a theory of \emph{sequences}
\cite{cvc5,z3,bjornerSMTLIBFormatSequences,DBLP:conf/cade/ShengNRZDGPQBT22}.
Sequences can be viewed as a generalization of strings and arrays: they are
indexed by integers, but their values do not have to be
characters.
\citet{DBLP:conf/smt/LamC14} provide an encoding of set comprehensions over linear integer arithmetic into linear integer arithmetic extended with an uninterpreted sort.
Syntax-guided synthesis (SyGuS) problems query solvers for functions satisfying
an input problem~\cite{DBLP:conf/sat/NotzliRBNPBT19}, and commonly support algebraic datatypes.
\citet{DBLP:conf/sat/BergerZNP0BT25} extend the theory of bit-vectors to handle symbolic widths.
Finally, some solvers, such as Rosette
\cite{torlakLightweightSymbolicVirtual2014}, only allow lists of symbolic elements if they have a concrete length.
With this approach, Rosette can lift Lisp-like concrete list operations (\eg,
\ml{cdr}, \ml{car}) to lists whose elements are symbolic elements, and thus benefit from functions defined in the underlying Racket language.
However, the user becomes responsible for finding suitable concrete lengths for symbolic lists.

\paragraph{Client applications of \duelist}
\duelist can be plugged into any SMT solver client requiring reasoning over lists.
One canonical application is symbolic execution; we refer the reader to the survey of~\citet{DBLP:journals/csur/BaldoniCDDF18}
for an in-breadth coverage.
\duelist is naturally suitable to improve concolic execution of Catala programs
\cite{DBLP:conf/esop/GoutagnyFM25}, which inspired both our running example
and several of our evaluation benchmarks.
In a similar fashion, \duelist could prove beneficial to symbolic execution
tools for functional programming languages~\cite{DBLP:journals/scp/GiantsiosPS17,DBLP:conf/pldi/HallahanXBJP19}.
More generally, we believe \duelist could be a helpful specialized solver for
larger deductive verification tasks~\cite{DBLP:conf/esop/FilliatreP13,DBLP:conf/popl/SwamyHKRDFBFSKZ16,DBLP:conf/ecoop/Porre0B23}.

\paragraph{Program analysis for data structures}
Our \fold list abstraction (\Cref{sec:fold:neutral}) is related to data structure summarization
techniques~\cite{kastner:inria-00528600,DBLP:conf/ecoop/ValnetMM25} from the static analysis community.
\citet{DBLP:conf/issta/PerryMZC17} introduce array transformations to speed up
symbolic execution within KLEE, using a methodology akin to array segmentation
\cite{DBLP:conf/popl/CousotCL11}.
\citet{DBLP:conf/sas/MonniauxA15} describe a generic, precision-tunable method to encode arrays and maps into purely scalar programs.
\citet{DBLP:conf/sas/MonniauxG16} encode array-manipulating programs into array-free Horn clauses, leading to automated functional correctness proof for some array programs.
Program analysis techniques mainly focus on analyzing
data structures, including linked lists,
through their low-level representation
\cite{DBLP:conf/popl/DilligDA11,DBLP:conf/pldi/BouajjaniDES11}, using ideas from
separation logic to create decision procedures~\cite{Reynolds2016} or conservative shape analyses
\cite{DBLP:journals/ftpl/ChangDMRR20}.
\duelist presents a different trade-off: it can be efficient by keeping the
representation of lists abstract and restricting the input language.

\section{Conclusion}%

Despite functional lists being a widely used data structure, they do not enjoy
first-class support within SMT solvers, therefore limiting the automation
of many program verification approaches.
To address this issue, this work introduced \duelist, an extension to off-the-shelf SMT solvers
performing an abstraction-refinement loop to solve input problems over abstract
lists with standard combinators (\len, \map, \fold).

\duelist relies on a cooperation between an over-approximating and an under-approximating
solver, which reason about list constraints through two notable classes of shape
abstractions; drawing inspiration from analyses developed for related data types,
such as arrays or strings, one focuses on length abstractions, while the other
investigates conditional \fold reductions.
Under the hood, \duelist leverages these abstractions to translate lists constraints
into list-free queries that can be discharged by state-of-the-art SMT solvers.

To evaluate our approach and foster future research on this topic,
we assembled a diverse corpus of size-parametric list benchmarks stemming
from previous works and real-world programs.
Our experimental evaluation demonstrates that \duelist extends the reasoning
capabilities of existing solvers, while outperforming
them in the vast majority of cases for previously supported list-related problems,
thus showcasing the benefits of our specialized theory and solving procedure.
In future work, we will focus on adding currently unsupported combinators to support pairs
of lists (e.g., \smtlib{fold2}), tackling technical limitations to support
lists of lists, and extending our prototype implementation to support a wider
range of SMT solvers through the use of SMT-agnostic libraries such as Smt.ml.

\bibliographystyle{ACM-Reference-Format}
\bibliography{conferences,cited,src}

\end{document}